\documentclass[reprint,superscriptaddress,amsmath,amssymb,prd,nofootinbib,longbibliography]{revtex4}
\usepackage{textcomp}
\usepackage[utf8]{inputenc}
\usepackage{amsmath}
\usepackage{amssymb}
\usepackage[bookmarks=false,
 breaklinks=false,pdfborder={0 0 1},
 colorlinks=false]
 {hyperref}
\hypersetup{
 bookmarks}

\makeatletter
\@ifundefined{textcolor}{}
{%
 \definecolor{BLACK}{gray}{0}
 \definecolor{WHITE}{gray}{1}
 \definecolor{RED}{rgb}{1,0,0}
 \definecolor{GREEN}{rgb}{0,1,0}
 \definecolor{BLUE}{rgb}{0,0,1}
 \definecolor{CYAN}{cmyk}{1,0,0,0}
 \definecolor{MAGENTA}{cmyk}{0,1,0,0}
 \definecolor{YELLOW}{cmyk}{0,0,1,0}
}

\@ifundefined{textcolor}{}{%
 \definecolor{BLACK}{gray}{0}
 \definecolor{WHITE}{gray}{1}
 \definecolor{RED}{rgb}{1,0,0}
 \definecolor{GREEN}{rgb}{0,1,0}
 \definecolor{BLUE}{rgb}{0,0,1}
 \definecolor{CYAN}{cmyk}{1,0,0,0}
 \definecolor{MAGENTA}{cmyk}{0,1,0,0}
 \definecolor{YELLOW}{cmyk}{0,0,1,0}
}

\usepackage{epsfig}\usepackage{mathrsfs}
\usepackage{mathrsfs}
\usepackage[usenames,dvipsnames]{color}
\usepackage[svgnames]{xcolor}
\usepackage{subfigure}
\usepackage{times}
\usepackage{yhmath}
\usepackage{braket}
\usepackage{soul}
\usepackage{comment}

\newcommand{\rr}{r}

\makeatother

\begin{document}

\title{General first-order constitutive equations for a relativistic dissipative multi-component fluid
in the presence of a weak background electromagnetic field}
\author{J. Félix Salazar}
\address{Departamento de Matemáticas Aplicadas y Sistemas, Universidad Autónoma
Metropolitana-Cuajimalpa (05348) Cuajimalpa de Morelos, Ciudad de
México México.}
\author{Ana Laura García-Perciante}
\address{Departamento de Matemáticas Aplicadas y Sistemas, Universidad Autónoma
Metropolitana-Cuajimalpa (05348) Cuajimalpa de Morelos, Ciudad de
México México.}
\author{Olivier Sarbach}
\address{Instituto de Física y Matemáticas, Universidad Michoacana de San Nicolás
de Hidalgo, Edificio C-3, Ciudad Universitaria, 58040 Morelia, Michoacán,
México.}

\begin{abstract}
In this work we establish general constitutive equations for a relativistic plasma composed of an arbitrary number of classical, charged, and chemically non-reacting species in the presence of a background electromagnetic field. The intensity of the electromagnetic force acting on either species is assumed to be comparable with the gradients of the state variables and is thus considered as a first-order driving force for dissipation within the gradient expansion. Assuming that the species share the same temperature and velocity flow when in equilibrium, we determine the entropy production and establish constitutive equations in terms of first-order frame invariant quantities. Conditions on the corresponding transport coefficients are given in order to comply with Onsager's reciprocal relations and the second law of thermodynamics. In particular, we discuss cross effects for the binary mixture which may be relevant for astrophysical and cosmological applications. Finally, we derive a complete set of evolution equations governing the dynamics of a binary mixture propagating on an arbitrary spacetime background.
\end{abstract}
\maketitle

\section{Introduction}

\label{Introduction}

Despite having been regarded as generically unstable and acausal for decades, first-order theories for relativistic dissipative fluids are currently being reconsidered. Recent works propose \textit{general first-order theories} in which pathological behavior can be avoided, provided suitable conditions on the state variables and the parameters  involved are met~\cite{Kovtun2019,BDN2018,BDN2019,Hoult2020,Bemfica2022,Disconzi2024,SGS2025A,SGS2025B,schianchi2026stronglyhyperbolicviscousrelativistic}. Therefore, these works offer physically sound models  for relativistic fluids in the sense that they lead to stable, causal, and hyperbolic systems of transport equations. These properties are achieved by considering non-equilibrium corrections to all components of the particle current and stress-energy tensor and allowing such deviations to be driven by first derivatives (both in space and time) of all state variables. Such theories are usually referred to as BDNK, named after their founding authors Bemfica, Disconzi, Noronha, and Kovtun. For the microscopic foundation of these theories, see~\cite{RochaDenicol21,Rocha2022,Kovtun2022,RDNR2024,JNET26b}.

The fact that the general first-order theories lead to physically sound dissipative hydrodynamic equations make them good candidates to describe non-equilibrium relativistic fluids. However, for most applications in high-energy physics, astrophysics, and cosmology, the system is composed of more than one species. For example, in the framework of Quantum Chromodynamics, strongly interacting matter under extreme conditions of temperature and density is expected to undergo a deconfinement transition to a state known as the quark-gluon plasma. In this state, quarks and gluons are no longer confined within hadrons \cite{Shuryak2017, Pasechnik2017,Ghiglieri2018,Jena2025}. A hydrodynamic description of the quark-gluon plasma has been performed, taking into account the system's multi-component nature. However, the analysis has been carried out using the Israel–Stewart theory and Grad's moment method \cite{Monnai2010, Monnai_2013}. 
In relativistic astrophysics, a detailed study of neutron stars including their superfluid core regions requires a multi-fluid description
 \cite{Chamel2008, Gusakov2013, Rau2020}. Further, the relevant components in neutron star mergers and their ejecta include baryonic matter (neutrons and protons), leptons (electrons and positrons), and neutrino radiation, which establishes a natural framework for a relativistic multifluid description \cite{MostHaber2024,ChabanovRezolla2025,ChabanovRezollaB,Fujibayashi2023,Just2015}. Also, accretion disks provide a natural setting for relativistic mixtures, as they involve interacting components such as ions, electrons, and radiation (photons or neutrinos), and cannot be reduced in general to a single effective fluid \cite{Yaun2014,Fernandez2022,Millauro2024}. Finally, in the cosmology realm, as the Universe expanded through its various epochs, the prevailing theory posits the coexistence of different kinds of materials such as radiation, dark energy, dark matter, and baryonic matter \cite{cMeB95,PlanckColl}. All these examples illustrate the natural need to generalize first-order dissipative fluid theories to the case of mixtures, where transport properties are modified and cross-effects become important.

In this work we present the natural extension of the general first-order theories for a single dissipative relativistic fluid to the case of an $\rr$-component non-reacting mixture, which we refer to simply as \textit{mixture} throughout the work. Relativistic mixtures have been previously addressed through Carter's theory \cite{Carter1989,Carter1992,Andersson2012,GavassinoMF2020}, transient thermodynamics \cite{Israel1976,Israel1979}, and relativistic kinetic theory \cite{Stewart-Book,DeGroot1980,CercignaniKremer-Book,Kremer2003} which also takes into account chemical reactions.  For other approaches, see for example Ref.~\cite{Dommes2020}. However, to the authors' knowledge, this type of systems have not been addressed within general first-order theories so far. It is important to clarify at the outset that in the present work, we assume that in equilibrium all species possess the same velocity vector $u^{\mu}$ and temperature $T$, such that the only state variables depending on the species are the particles' number densities $n_i$.\footnote{This assumption is supported by a kinetic theory study of a collisional gas mixture where it can be shown that zero entropy production leads to species sharing the same hydrodynamic velocity and temperature (see Refs.~\cite{DeGroot1980,CercignaniKremer-Book,JNET26c} for the details of such calculation).} Also, we assume the electromagnetic field to be external such that fields generated by local currents or total charge densities are neglected. Moreover, we assume that the background electromagnetic field is weak in the sense that the electromagnetic force acting on all components of the fluid scales as the gradients of the state variables and is thus here considered as a first-order driving force for dissipative fluxes.

This article is organized as follows. In section~\ref{Section: Constitutive Relations}, the most general first-order constitutive equations for the mixture are presented. In particular, we explicitly construct transport coefficients which are invariant under a change of frame and representation, and relevant examples of frames are briefly discussed. In~section~\ref{Section: Entropy Current} an expression for the entropy production is established, which is form-invariant under a change of frame. Based on this expression we discuss the restrictions that arise when imposing Onsager's reciprocal relations and the second law of thermodynamics. Next, in section~\ref{Sec: Binary Mixture} we restrict our analysis to the case of a binary mixture. We explicitly identify the transport coefficients with those known in the literature and analyze the cross effects by examining the phenomenology in a particular frame. In section~\ref{Sec:Binary Mixture in the TFP-Frame}, the complete set of evolution equations that govern our fluid theory in a trace-fixed particle frame for the binary mixture is presented. Conclusions are drawn in section~\ref{Section:Conclusions}. Finally, appendix~\ref{App:Thermo} reviews some thermodynamic relations that are relevant to this article.

We assume the spacetime background is described by a $C^\infty$-smooth, $(d+1)$-dimensional Lorentzian manifold $(M,g)$ which is globally hyperbolic, time-oriented, and whose metric has signature $(-,+,\ldots,+)$. We adopt units in which the speed of light is one. Greek indices denote spacetime indices and run over $0,1,\ldots,d$, while Latin indices label different species, and run over $1,2,...,\rr$ with $\rr$ arbitrary. $\Delta_{\mu}{}^{\nu}:=\delta_{\mu}{}^{\nu}+u_{\mu}u^{\nu}$ denotes the projector orthogonal to the velocity vector $u^{\mu}$. Symmetrization and anti-symmetrization of the indices is denoted by
$A^{(\mu\nu)}:=\frac{1}{2}\left(A^{\mu\nu}+A^{\nu\mu}\right)$ and 
$A^{[\mu\nu]}:=\frac{1}{2}\left(A^{\mu\nu}-A^{\nu\mu}\right)$, respectively. Finally, $\nabla_{\mu}$ refers to the covariant derivative associated with $g$, and the operators $D_\mu$ and $\dot{\left(\,\right)}$ to the spatial covariant derivative and derivative along $u^{\mu}$, respectively, which for scalar quantities are defined as $D_{\mu}:=\Delta_{\mu}^{\ \nu}\nabla_{\nu}$ and $\dot{\left(\,\right)}:=u^{\mu}\nabla_{\mu}$. For the definition of these operators when acting on an arbitrary rank tensor field orthogonal to $u^\mu$, see Appendix~C of Ref.~\cite{SGS2025B}.

\section{Constitutive relations and invariants for a mixture}
\label{Section: Constitutive Relations} 

As outlined in Section~\ref{Introduction}, the aim of this article is to extend general first-order theories to the case of mixtures. To this end, we consider a system described by $\rr$ particle currents $J_{i}^{\mu}$ and a stress--energy tensor $T^{\mu\nu}= T^{\nu\mu}$, which satisfy the following balance laws:
\begin{subequations}
\label{Eq:BalanceLaws}
\begin{eqnarray}
\nabla_{\mu}J_{i}^{\mu} &=& 0,
\qquad i=1,2,\ldots,\rr,
\label{Eq:BalanceLaws1}\\
\nabla_{\mu}T^{\mu\nu}+I_{\mu}F^{\mu\nu} &=& 0,
\label{Eq:BalanceLaws2}
\end{eqnarray}
\end{subequations}
where 
\begin{equation}\label{Eq:CurrentI}
I^{\mu}:=\sum_{i=1}^{\rr}q_{i}J_{i}^{\mu},
\end{equation}
denotes the electric current vector with $q_{i}$ the electric charge of the $i-$th species and the $2$-form $F^{\mu \nu}$ denotes the external electromagnetic field. Performing a $d+1$ decomposition, the particle currents and stress-energy tensor can be written in general as:
\begin{subequations}
\label{Eq:JTKovtun}
\begin{align}
J_{i}^{\mu} & =\mathcal{N}_{i}u^{\mu}+\mathcal{J}_{i}^{\mu},
\qquad\qquad i=1,2,\ldots,\rr,
\label{Eq:JmunuKovtun}\\
T^{\mu\nu} & =\mathcal{E}u^{\mu}u^{\nu}+\mathcal{P}\Delta^{\mu\nu}+2u^{(\mu}\mathcal{Q}^{\nu)}+\mathcal{T}^{\mu\nu},
\label{Eq:TmunuKovtun}
\end{align}
\end{subequations}
with $\mathcal{J}_{1,2,...,\rr}^{\mu}$, $\mathcal{Q}^{\mu}$, and $\mathcal{T}^{\mu\nu}$ being orthogonal to the future-directed timelike unit velocity vector $u^{\mu}$, and additionally $\mathcal{T}^{\mu\nu}$ being symmetric and trace-free. Notice that in Eq.~(\ref{Eq:BalanceLaws1}) the particle number density is conserved for each species, which is the consequence of the absence of chemically reactions. In contrast, only the total stress--energy tensor appears in the energy and momentum balance~(\ref{Eq:BalanceLaws2}), which means that energy and momentum
can be interchanged between species.

When the system is in equilibrium (perfect multi-fluid case) Eqs.~(\ref{Eq:JTKovtun}) reduce to: 
\begin{equation}
J_{i}^{\mu}=n_{i}u^{\mu},\quad T^{\mu\nu}=neu^{\mu}u^{\nu}+p\Delta^{\mu\nu}.\label{eq:II.3}
\end{equation}
Here $n_{i}$ is $i-$th particle number density and $n:=\sum\limits_{i=1}^{\rr}n_{i}$. Also, $p:=nk_{B}T$ and $e:=\frac{1}{n}\sum\limits_{i=1}^{\rr}n_{i}e_{i}(T)$ with $e_{i}(T)$ being the internal energy per particle for the $i-$th species and $T$ the temperature. In this scenario, the balance equations~(\ref{Eq:BalanceLaws}) correspond to the relativistic Euler equations for a charged mixture in the presence of an external electromagnetic field, which are expressed as follows:
\begin{subequations}
\begin{align}
 & \dot{n}_{i}+n_{i}\theta=0,\label{eq:II.4}\\
 & a_{\mu}+\frac{1}{nh}D_{\mu}p-\frac{q}{h}E_{\mu}=0,\label{eq:II.5}\\
 & \frac{\dot{T}}{T}+\frac{k_{B}}{c_{v}}\theta=0.\label{eq:II.6}
\end{align}
\end{subequations}
Here the acceleration is defined as $a_{\mu}:=\dot{u}_{\mu}=u^{\nu}\nabla_{\nu}u_{\mu}$
and $\theta:=\nabla_{\mu}u^{\mu}$ stands for the expansion. The specific heat at constant volume $c_{v}$ and enthalpy $h$ are defined as
$c_{v}:=\left.\frac{\partial e}  {\partial T}\right|_{n_1,n_2,\ldots,n_r}$ and $h:=\frac{1}{n}\sum\limits_{i=1}^{\rr}n_{i}h_{i}$, where $h_i:=e_i+k_{B}T$ and thus $h=e+k_{B}T$. Also, we have introduced the electric field $E^{\mu}:=F^{\mu\nu}u_{\nu}$ measured by co-moving observers  and the local net charge density $nq:=\sum\limits_{i=1}^{\rr}n_{i}q_{i}$.

Out of equilibrium, $J_{i}^{\mu}$ and $T^{\mu\nu}$ have the general expressions given in Eqs.~(\ref{Eq:JTKovtun}) where the calligraphic quantities include
non-equilibrium contributions. The balance equations~\eqref{Eq:BalanceLaws} yield
\begin{subequations}
\label{Eq:Balance}
\begin{align}
 & \dot{\mathcal{N}_{i}}+\mathcal{N}_{i}\theta+\left(D_{\mu}+a_{\mu}\right)\mathcal{J}_{i}^{\mu}=0,\label{eq:ndotcal}\\
 & \dot{\mathcal{E}}+\left(\mathcal{E}+\mathcal{P}\right)\theta+\left(D_{\mu}+2a_{\mu}\right)\mathcal{Q}^{\mu}+\mathcal{T}^{\mu\nu}\sigma_{\mu\nu}-I_{\mu}E^{\mu}=0,\label{eq:edotcal}\\
 & \left(\mathcal{E}+\mathcal{P}\right)a^{\alpha}+D^{\alpha}\mathcal{P}+\left(D_{\mu}+a_{\mu}\right)\mathcal{T^{\mu\alpha}}+\frac{d+1}{d}Q^{\alpha}\theta+\dot{\mathcal{Q}}^{\alpha}+\mathcal{Q}_{\mu}\left(\sigma^{\mu\alpha}+\omega^{\mu\alpha}\right)+\Delta_{\ \nu}^{\alpha}I_{\mu}F^{\mu\nu}=0,\label{eq:udotcal}
\end{align}
\end{subequations}
where $\sigma_{\mu\nu}:=\left(\Delta_{(\mu}{}^{\alpha}\Delta_{\nu)}{}^{\beta}-d^{-1}\Delta_{\mu\nu}\Delta^{\alpha\beta}\right)\nabla_{\alpha}u_{\beta}$
and $\omega_{\mu\nu}:=\Delta_{[\mu}{}^{\alpha}\Delta_{\nu]}{}^{\beta}\nabla_{\alpha}u_{\beta}$
are the traceless symmetric and antisymmetric parts of $\nabla_{\alpha}u_{\beta}$, which constitute the shear and vorticity associated with $u_{\mu}$.

The general first-order theory states that the dissipative contributions to the calligraphic quantities in Eqs.~(\ref{Eq:JTKovtun}) and (\ref{Eq:Balance}), are driven by the gradients of the state variables to first order. The ensuing relations between dissipative quantities and their driving forces are referred to as constitutive equations. Thus, assuming that the system is near thermodynamic equilibrium, these relations can be expressed as a Taylor series expansion to first-order in the derivatives of the state variables $(n_{1},..,n_{\rr},T,u^{\mu})$ with the addition of an electromagnetic force term which, as pointed out above, is here assumed to be a first-order quantity in the expansion (see Ref.~\cite{SGS2025B} for the analog proposal in the single fluid case). Adopting the same notation as in Ref.~\cite{SGS2025B} for the sake of consistency and ease of comparison (see also Refs.~\cite{Kovtun_2012,Kovtun2019}), the first-order general constitutive equations for the mixture are written as:
\begin{subequations}
\label{Eq:Constitutive}
\begin{align}
\mathcal{N}_{i} & =n_{i}+\sum\limits_{j=1}^{\rr}\nu_{1ij}\frac{\dot{n}_{j}}{n}+\nu_{2i}\frac{\dot{T}}{T}+\nu_{3i}\theta+\mathcal{O}(\partial^{2}),
\label{Eq:ConstitutiveN}\\
\mathcal{E} & =ne+\sum\limits_{j=1}^{\rr}\varepsilon_{1j}\frac{\dot{n}_{j}}{n}+\varepsilon_{2}\frac{\dot{T}}{T}+\varepsilon_{3}\theta+\mathcal{O}(\partial^{2}),
\label{Eq:ConstitutiveE}\\
\mathcal{P} & =nk_{B}T+\sum\limits_{j=1}^{\rr}\pi_{1j}\frac{\dot{n}_{j}}{n}+\pi_{2}\frac{\dot{T}}{T}+\pi_{3}\theta+\mathcal{O}(\partial^{2}),
\label{Eq:ConstitutiveP}\\
\mathcal{J}_{i}^{\mu} & =\sum\limits_{j=1}^{\rr}\gamma_{1ij}\frac{D^{\mu}n_{j}}{n}+\gamma_{2i}\frac{D^{\mu}T}{T}+\gamma_{3i}a^{\mu}+\gamma_{4i}\frac{q}{h}E^{\mu}+\mathcal{O}(\partial^{2}),
\label{Eq:ConstitutiveJ}\\
\mathcal{Q}^{\mu} & =\sum\limits_{j=1}^{\rr}\kappa_{1j}\frac{D^{\mu}n_{j}}{n}+\kappa_{2}\frac{D^{\mu}T}{T}+\kappa_{3}a^{\mu}+\kappa_{4}\frac{q}{h}E^{\mu}+\mathcal{O}(\partial^{2}),
\label{Eq:ConstitutiveQ}\\
\mathcal{T}^{\mu\nu} & =-2\eta\sigma^{\mu\nu}+\mathcal{O}(\partial^{2}).
\label{Eq:ConstitutiveT}
\end{align}
\end{subequations}
We stress that the underlying justification for the general expressions in Eqs.~(\ref{Eq:Constitutive}) is based on the assumption that deviations from thermal equilibrium are small and can be expressed in terms of first-order gradients of the state variables. It is important to realize that this expansion is not unique, since it is subject to two types of freedom. The first, referred to as the choice of frame~\cite{Kovtun_2012,Kovtun2019} arises from the fact that, out of equilibrium, there is no unique way to define the state variables $(n_{1},..,n_{\rr},T,u^{\mu})$. Any choice for these variables is as valid as any other, provided they reduce to the correct expressions when in equilibrium. The second freedom, referred to as the choice of representation~\cite{JNET24,JNET26b,JNET26a}, arises from the observation that the expansions in Eqs.~(\ref{Eq:Constitutive}) imply that
\begin{subequations}
\label{Eq:EulerTrunc}
\begin{align}
\frac{\dot{n}_{i}}{n_{i}}+\theta=\mathcal{O}(\partial^{2}),
\label{Eq:EulerTrunc1-1}\\
\frac{\dot{T}}{T}+\frac{k_{B}}{c_{v}}\theta=\mathcal{O}(\partial^{2}),
\label{Eq:EulerTrunc2-1}\\
a^{\mu}+\frac{k_{B}T}{h}\left(\frac{D^{\mu}n}{n}+\frac{D^{\mu}T}{T}\right)-\frac{q}{h}E^{\mu}=\mathcal{O}(\partial^{2}),
\label{Eq:EulerTrunc3-1}
\end{align}
\end{subequations}
such that one can add multiples of the combinations above to the first-order constitutive equations without altering them to this order. This second freedom allows one to consider, in principle, couplings between all possible fluxes and forces (gradients of the state variables and the electromagnetic force) of the same rank. Therefore, the general expressions in Eqs.~\eqref{Eq:Constitutive} are completely justified in the sense that all the proposed terms may be present. However, it is clear that not all the quantities involved in them are of physical nature. One needs to identify the coefficients (or, more precisely, combinations of them) that remain invariant under the two transformations described above and write the theory in terms of such invariant combinations in order to draw physically meaningful conclusions. The rest of this section is devoted to the construction of such invariant coefficients, together with frame-independent fluxes.\footnote{Since the derivatives and the electromagnetic force terms on the right-hand sides of Eqs.~(\ref{Eq:Constitutive}) are already first-order quantities, they are invariant under both transformations and hence only the invariance of the coefficients and the fluxes are addressed.}

\subsection{First-order frame transformations}\label{SubSec:First-order frame transformations}

In order to analyze the effects of a first-order change of frame, we consider the following transformation of the state variables:
\begin{subequations}
\label{Eq:ChangeOfFrame}
\begin{eqnarray}
n_{i} & \mapsto & n_{i}':=n_{i}+\delta n_{i},\quad i=1,2,\ldots,\rr,\label{Eq:ChangeOfFramen}\\
T & \mapsto & T':=T+\delta T,\label{Eq:ChangeOfFrameT}\\
u^{\mu} & \mapsto & u'{}^{\mu}:=u^{\mu}+\delta u^{\mu},\label{Eq:ChangeOfFrameu}
\end{eqnarray}
\end{subequations}
where $\delta n_{i}$, $\delta T$, and $\delta u^{\mu}$ are of order $\mathcal{O}(\partial)$, and the normalization condition required for
the velocity $u'{}^{\mu}u'_{\mu}=-1$, implies $u^{\mu}\delta u_{\mu}=\mathcal{O}(\partial^{2})$. Consequently, the components of $J_{i}^{\mu}$ and $T^{\mu\nu}$ in the new frame are related to those in the original one in the following manner:
\begin{subequations}
\begin{eqnarray}
\mathcal{N}'_{i} & = & \mathcal{N}_{i}+\mathcal{O}(\partial^{2}),\label{Eq:InvariantNa}\\
\mathcal{E}' & = & \mathcal{E}+\mathcal{O}(\partial^{2}),\label{Eq:InvariantE}\\
\mathcal{P}' & = & \mathcal{P}+\mathcal{O}(\partial^{2}),\label{Eq:InvariantP}\\
\mathcal{J}'{}_{i}^{\mu} & = & \mathcal{J}_{i}^{\mu}-n_{i}\delta u^{\mu}+\mathcal{O}(\partial^{2}),\label{Eq:InvariantJ}\\
\mathcal{Q}'{}^{\mu} & = & \mathcal{Q}^{\mu}-nh\delta u^{\mu}+\mathcal{O}(\partial^{2}),\label{Eq:InvariantQ}\\
\mathcal{T}'{}^{\mu\nu} & = & \mathcal{T}^{\mu\nu}+\mathcal{O}(\partial^{2}).\label{Eq:InvariantT}
\end{eqnarray}
\end{subequations}
Additionally, the $\mathcal{O}(\partial)$ terms in Eqs.~\eqref{Eq:ChangeOfFrame} can be expanded according to
\begin{subequations}
\label{Eq:deltatransf}
\begin{align}
\delta n_{i} & =\sum\limits_{j=1}^{\rr}\alpha_{1ij}\frac{\dot{n}_{j}}{n}+\alpha_{2i}\frac{\dot{T}}{T}+\alpha_{3i}\theta+\mathcal{O}(\partial^{2}),\\
\delta T & =\sum\limits_{j=1}^{\rr}\beta_{1j}\frac{\dot{n}_{j}}{n}+\beta_{2}\frac{\dot{T}}{T}+\beta_{3}\theta+\mathcal{O}(\partial^{2}),\\
\delta u^{\mu} & =\sum\limits_{j=1}^{\rr}\mu_{1j}\frac{D^{\mu}n_{j}}{n}+\mu_{2}\frac{D^{\mu}T}{T}+\mu_{3}a^{\mu}+\mu_{4}\frac{q}{h}E^{\mu}+\mathcal{O}(\partial^{2}),
\end{align}
\end{subequations}
where the quantities $\alpha_{1ij},\alpha_{Ii},\beta_{1j},\beta_{I},\mu_{1j},\mu_{J}$ for $I=2,3$ and $J=2,3,4$, can only depend on $(n_{1},...,n_{\rr},T)$. Considering these hypotheses, the coefficients in the new frame can be shown to be given by $\eta'=\eta$ and
\begin{subequations}
\label{Eq:FrameTransform}
\begin{align}
\nu'_{1ij} & =\nu_{1ij}-\alpha_{1ij},\qquad\nu'_{Ii}=\nu_{Ii}-\alpha_{Ii},
\label{Eq:Tranus}\\
\varepsilon'_{1j} & =\varepsilon_{1j}-\sum\limits_{i=1}^{\rr}\alpha_{1ij}e_{i}-nc_{v}\beta_{1j},\quad\varepsilon'_{I}=\varepsilon_{I}-\sum\limits_{i=1}^{\rr}\alpha_{Ii}e_{i}-nc_{v}\beta_{I},
\label{Eq:Traepsilon}\\
\pi'_{1j} & =\pi_{1j}-k_{B}T\sum\limits_{i=1}^{\rr}\alpha_{1ij}-nk_{B}\beta_{1j},\quad\pi'_{I}=\pi_{I}-k_{B}T\sum\limits_{i=1}^{\rr}\alpha_{Ii}-nk_{B}\beta_{I},
\label{Eq:Trapis}\\
\gamma'_{1ij} & =\gamma_{1ij}-n_{i}\mu_{1j},\quad\gamma'_{Ji}=\gamma_{Ji}-n_{i}\mu_{J},
\label{Eq:Tragammas}\\
\kappa'_{1j} & =\kappa_{1j}-nh\mu_{1j},\quad\kappa'_{J}=\kappa_{J}-nh\mu_{J},
\label{Eq:Trakappas}
\end{align}
\end{subequations}
where $I=2,3$ and $J=2,3,4$. Notice that while $\eta$ is already frame independent, this property does not extend to the remaining coefficients. Nevertheless, it is possible to construct the following linear combinations which are invariant under the transformations given by Eqs.~\eqref{Eq:FrameTransform}: 
\begin{subequations}
\label{Eq:flDef}
\begin{align}
f_{1j} & :=\pi_{1j}-\frac{k_{B}}{c_{v}}\varepsilon_{1j}+\sum\limits_{i=1}^{\rr}\left[\frac{k_{B}}{c_{v}}e_{i}-k_{B}T\right]\nu_{1ij},
\label{eq:f1b}\\
f_{I} & :=\pi_{I}-\frac{k_{B}}{c_{v}}\varepsilon_{I}+\sum\limits_{i=1}^{\rr}\left[\frac{k_{B}}{c_{v}}e_{i}-k_{B}T\right]\nu_{Ii},\qquad I=2,3,
\label{eq:fjb}\\
\ell_{1ij} & :=\gamma_{1ij}-\frac{1}{h}\frac{n_{i}}{n}\kappa_{1j},\quad\ell_{Ji}:=\gamma_{Ji}-\frac{1}{h}\frac{n_{i}}{n}\kappa_{J},\qquad J=2,3,4.
\label{eq:lki}
\end{align}
\end{subequations}
These invariants reduce to the ones given in Eqs.~(A.28) and (A.29) in Ref.~\cite{SGS2025B} for the single species case. 

We can now establish constitutive equations that hold in any frame by expressing them in terms of the coefficients defined in Eqs.~(\ref{Eq:flDef}). To this end, firstly, we construct a frame-invariant scalar non-equilibrium flux. It should be noted that, even though the quantities $\mathcal{N}_{1,2,\cdots,\rr}$, $\mathcal{E}$, ane $\mathcal{P}$ are frame invariant {[}see Eqs.~(\ref{Eq:InvariantNa} - \ref{Eq:InvariantP}){]} the non-equilibrium  contributions $\left(\mathcal{E}-ne\right)$, $(\mathcal{P}-p)$ and $\left(\mathcal{N}_{i}-n_{i}\right)$ are not, and in fact transform according to:
\begin{subequations}
\begin{align}
\mathcal{E}'-(ne)' & =\mathcal{E}-ne-\sum\limits_{i=1}^{\rr}e_{i}\delta n_{i}-nc_{v}\delta T+\mathcal{O}(\partial^{2}),\\
\mathcal{P}'-p' & =\mathcal{P}-p-k_{B}\left(n\delta T+T\delta n\right)+\mathcal{O}(\partial^{2}),\\
\mathcal{N}'_{i}-n'_{i} & =\mathcal{N}_{i}-n_{i}-\delta n_{i}+\mathcal{O}(\partial^{2}), \hspace{2cm} i=1,2,\dots,\rr,
\end{align}
\end{subequations}
with $\delta n:=\sum\limits_{i=1}^{\rr}\delta n_{i}$. However, the following combination is a first-order frame-invariant quantity: 
\begin{equation}
\hat{\mathrm{P}}:=-\frac{k_{B}}{c_{v}}\left[\left(\mathcal{E}-ne\right)-\frac{c_{v}}{k_{B}}\left(\mathcal{P}-p\right)-\sum\limits_{i=1}^{\rr}\left(e_{i}-c_{v}T\right)\left(\mathcal{N}_{i}-n_{i}\right)\right].\label{Eq:Scalar Invariant-1}
\end{equation}
Using Eqs.~\eqref{Eq:ConstitutiveN},\eqref{Eq:ConstitutiveE}, and \eqref{Eq:ConstitutiveP}, this can be expressed in terms of the invariant combinations given in Eqs.~\eqref{eq:f1b} and \eqref{eq:fjb} as follows:
\begin{equation}
\hat{\mathrm{P}}=\sum\limits_{j=1}^{\rr}f_{1j}\frac{\dot{n}_{j}}{n}+f_{2}\frac{\dot{T}}{T}+f_{3}\theta+\mathcal{O}(\partial^{2}).\label{Eq:Scalar Invariantfs}
\end{equation}
Like the scalar fluxes $\mathcal{E}-ne$, $\mathcal{P}-p$, and $\mathcal{N}_{i}-n_{i}$, the vector fluxes are not frame invariant {[}see Eqs.~(\ref{Eq:InvariantJ}) and (\ref{Eq:InvariantQ}){]}. However, one can define the following combinations
\begin{subequations}
\label{Eq:VectorInvariants}
\begin{align}
\hat{\mathcal{J}_{i}^{\mu}} & :=\mathcal{J}_{i}^{\mu}-\frac{n_{i}}{n}\sum\limits_{j=1}^{\rr}\mathcal{J}_{j}^{\mu},\qquad i=1,2,\dots,\rr,
\label{Eq:Vector Invariant J-1}\\
\hat{Q}^{\mu} & :=Q^{\mu}-\sum\limits_{i=1}^{\rr}h_{i}\mathcal{J}_{i}^{\mu},
\label{Eq:Vector Invariant Q-1}
\end{align}
\end{subequations}
which remain unaltered when a first-order change of frame is performed. Notice that $\sum\limits_{i=1}^{\rr}\hat{\mathcal{J}_{i}^{\mu}}=0$, and thus one has only $(\rr - 1)$ independent invariant fluxes $\hat{\mathcal{J}_{i}^{\mu}}$. Using Eqs.~(\ref{Eq:ConstitutiveJ}) and (\ref{Eq:ConstitutiveQ}), together with Eq.~(\ref{eq:lki}), the vector constitutive equations can be written in a frame-invariant manner as follows:
\begin{subequations}
\label{Eq:JQhat}
\begin{align}
\hat{\mathcal{J}}_{i}^{\mu} & =\sum_{j=1}^{\rr}\hat{\ell}_{1ij}\frac{D^{\mu}n_{j}}{n}+\hat{\ell}_{2i}\frac{D^{\mu}T}{T}+\hat{\ell}_{3i}a^{\mu}+\hat{\ell}_{4i}\frac{q}{h}E^{\mu}+\mathcal{O}(\partial^{2}),\qquad i=1,2,\cdots,\rr,
\label{eq:jhat}\\
\hat{\mathcal{Q}}^{\mu} & =-\sum_{i,j=1}^{\rr}h_{i}\ell_{1ij}\frac{D^{\mu}n_{j}}{n}-\sum_{i=1}^{\rr}h_{i}\ell_{2i}\frac{D^{\mu}T}{T}-\sum_{i=1}^{\rr}h_{i}\ell_{3i}a^{\mu}-\sum_{i=1}^{\rr}h_{i}\ell_{4i}\frac{q}{h}E^{\mu}+\mathcal{O}(\partial^{2}),
\label{eq:qhat}
\end{align}
\end{subequations}
where $\hat{\ell}_{1ij}:=\ell_{1ij}-\dfrac{n_{i}}{n}\sum\limits_{k=1}^{\rr}\ell_{1kj}$ and $\hat{\ell}_{Ji}:=\ell_{Ji}-\dfrac{n_{i}}{n}\sum\limits_{j=1}^{\rr}\ell_{Jj}$
for $J=2,3,4$.

It is important to emphasize at this point that the definitions of the invariant fluxes, namely $\hat{\mathcal{J}}_{i}^{\mu}$ and $\hat{\mathcal{Q}}^{\mu}$ given by Eqs.~(\ref{Eq:Vector Invariant J-1}) and (\ref{Eq:Vector Invariant Q-1}) coincide precisely with the expressions for the diffusion flux and the reduced heat flow introduced previously in the literature in a frame for which $\mathcal{N}_{i}=n_{i}$ (see Eq.~(24) on page 26 in~\cite{GrootMazur1983} and Sections I.1.g and V.2.c in~\cite{DeGroot1980}). This reduced heat flow is defined by subtracting from the heat flow the enthalpy transported by diffusing species which leads to a flow containing only thermal dissipation. Interestingly, we arrived at the same expressions for the diffusion and reduced heat flows as in Ref.~\cite{DeGroot1980} by looking for frame-invariant quantities, which reinforces their physical relevance.

For the scalar part, the definition of the invariant scalar flux given by Eq.~\eqref{Eq:Scalar Invariant-1} has no non-relativistic counterpart. This is due to the fact that the non-relativistic theory only considers changes of reference frames while enforcing $\mathcal{N}_{i}=n_{i}$ and $\mathcal{E}=ne$. Therefore, the only possible out-of-equilibrium scalar contribution is given by $\mathcal{P}-p$. In the present case however, since changes in the particle density and temperature are also considered, $\mathcal{E}$ and $\mathcal{N}_{i}$ also feature dissipative effects which leads to the construction of the general scalar invariant $\hat{\mathrm{P}}$. Notice that standard textbooks on relativistic kinetic theory (e.g.~\cite{DeGroot1980,CercignaniKremer-Book}) also restrict their analysis to frames for which the particle densities and the internal energy of the fluid are purely equilibrium quantities, such that only the pressure deviator appears in the scalar sector. 

\subsection{First-order changes of representation}
\label{subsec:changeofrepresentation}

As discussed above, one has the freedom to add multiples of the second-order combinations in Eqs.~(\ref{Eq:EulerTrunc}) to Eqs.~(\ref{Eq:Constitutive}) without altering the latter within the first-order theory. This leads to the following changes in the coefficients 
\begin{align}
 & \begin{aligned}\hat{\nu}_{1ij} & =\nu_{1ij}+\frac{n}{n_{j}}\Omega_{\nu ij}, & \hat{\nu}_{2i} & =\nu_{2i}+\chi_{\nu1}, & \hat{\nu}_{3i} & =\nu_{3i}+\sum_{j=1}^{\rr}\Omega_{\nu ij}+\frac{k_{B}}{c_{v}}\chi_{\nu i},\\
\hat{\varepsilon}_{1j} & =\varepsilon_{1j}+\frac{n}{n_{j}}\Omega_{\varepsilon j}, & \hat{\varepsilon}_{2} & =\varepsilon_{2}+\chi_{\varepsilon}, & \hat{\varepsilon}_{3} & =\varepsilon_{3}+\sum_{j=1}^{\rr}\Omega_{\varepsilon j}+\frac{k_{B}}{c_{v}}\chi_{\varepsilon},\\
\hat{\pi}_{1j} & =\pi_{1j}+\frac{n}{n_{j}}\Omega_{\pi j}, & \hat{\pi}_{2} & =\pi_{2}+\chi_{\pi}, & \hat{\pi}_{3} & =\pi_{3}+\sum_{j=1}^{\rr}\Omega_{\pi j}+\frac{k_{B}}{c_{v}}\chi_{\pi},\\
\hat{\gamma}_{1ij} & =\gamma_{1ij}+\frac{k_{B}T}{h}\xi_{\mathcal{J}i}, & \hat{\gamma}_{2i} & =\gamma_{2i}+\frac{k_{B}T}{h}\xi_{\mathcal{J}i}, & \hat{\gamma}_{3i} & =\gamma_{3i}+\xi_{\mathcal{J}i}, & \hat{\gamma}_{4i} & =\gamma_{4i}-\xi_{\mathcal{J}i},\\
\hat{\kappa}_{1j} & =\kappa_{1j}+\frac{k_{B}T}{h}\xi_{Q}, & \hat{\kappa}_{2} & =\kappa_{2}+\frac{k_{B}T}{h}\xi_{Q}, & \hat{\kappa}_{3} & =\kappa_{3}+\xi_{Q}, & \hat{\kappa}_{4} & =\kappa_{4}-\xi_{Q},
\end{aligned}
\end{align}
for $i,j=1,2,\cdots,\rr$, which is the result of adding the terms $\sum\limits_{j=1}^{\rr}\Omega_{\nu ij}\left(\frac{\dot{n}_{j}}{n_{j}}+\theta\right)$ and $\chi_{\nu i}\left(\frac{\dot{T}}{T}+\frac{k_{B}}{c_{v}}\theta\right)$ to $\mathcal{N}_{i}$ in Eq.~\eqref{Eq:ConstitutiveN}, and similarly
with Eqs.~\eqref{Eq:ConstitutiveE} and \eqref{Eq:ConstitutiveP} changing $\nu$ by either $\varepsilon$ or $\pi$ in the subscript of the coefficients. For the vector fluxes, we added $\xi_{\mathcal{J}i}\left(a^{\mu}+\dfrac{1}{nh}D^{\mu}p-\frac{q}{h}E^{\mu}\right)$ to $\mathcal{J}_{i}$ in Eq.~\eqref{Eq:ConstitutiveJ} and the same combination multiplied by $\xi_{Q}$ to Eq.~\eqref{Eq:ConstitutiveQ}. Since there are no second-order tensor combinations, the equation for $\mathcal{T}^{\mu\nu}$ remains unaltered under a change of representation. The following combinations are invariant with respect to these changes:
\begin{align}
 & \begin{aligned} & \sum_{j=1}^{\rr}\frac{n_{j}}{n}\nu_{1ij}+\frac{k_{B}}{c_{v}}\nu_{2i}-\nu_{3i}, &  & \sum_{j=1}^{\rr}\frac{n_{j}}{n}\varepsilon_{1j}+\frac{k_{B}}{c_{v}}\varepsilon_{2}-\varepsilon_{3},\quad\sum_{j=1}^{\rr}\frac{n_{j}}{n}\pi_{1j}+\frac{k_{B}}{c_{v}}\pi_{2}-\pi_{3},\\
 & \gamma_{2i}-\frac{k_{B}T}{h}\gamma_{3i},\quad\gamma_{3i}+\gamma_{4i}, &  & \gamma_{1ij}-\frac{k_{B}T}{h}\gamma_{3i},\quad\kappa_{2}-\frac{k_{B}T}{h}\kappa_{3},\quad\kappa_{1j}-\frac{k_{B}T}{h}\kappa_{3},\quad\kappa_{3}+\kappa_{4}.
\end{aligned}
\end{align}
When combined with Eqs.~(\ref{Eq:flDef}) they lead to the frame- and representation-invariant quantities, given by
\begin{subequations}
\label{Eq:Invariants}
\begin{align}
\eta,\label{Eq:Invariant eta}\\
\zeta & :=\sum_{j=1}^{\rr}\frac{n_{j}}{n}f_{1j}+f_{2}\frac{k_{B}}{c_{v}}-f_{3},\label{Eq:Invariant zeta}\\
\lambda_{1ij} & :=\ell_{1ij}-\frac{k_{B}T}{h}\ell_{3i},\label{Eq:InvariantL1}\\
\lambda_{2i} & :=\ell_{2i}-\frac{k_{B}T}{h}\ell_{3i}.\label{Eq:InvariantL2}\\
\lambda_{4i} & :=\ell_{4i}+\ell_{3i}.\label{Eq:InvariantL4}
\end{align}
\end{subequations}
Notice that the number of independent invariant coefficients so far is $\rr^{2}+2\rr+2$. Having reached this point, it is now possible to rewrite the scalar and vector frame-invariant fluxes in Eqs.~\eqref{Eq:Scalar Invariantfs} and \eqref{Eq:JQhat} in terms of these coefficients plus terms that depend solely on the choice of representation:
\begin{subequations}
\begin{align}
\hat{\mathrm{P}} & =-\zeta\theta+\sum_{j=1}^{\rr}\frac{n_{j}}{n}f_{1j}\left(\frac{\dot{n}_{j}}{n_{j}}+\theta\right)+f_{2}\left(\frac{\dot{T}}{T}+\frac{k_{B}}{c_{v}}\theta\right),
\label{Eq:Phat}\\
\hat{\mathcal{J}}_{i}^{\mu} & =\sum_{j=1}^{\rr}\hat{\lambda}_{1ij}\frac{D^{\mu}n_{j}}{n}+\hat{\lambda}_{2i}\frac{D^{\mu}T}{T}+\hat{\lambda}_{4i}\frac{q}{h}E^{\mu}+\hat{\ell}_{3i}\left(a^{\mu}+\frac{1}{nh}D^{\mu}p-\frac{q}{h}E^{\mu}\right),
\label{Eq:hatJ1-A}\\
\hat{\mathcal{Q}}^{\mu} & =-\sum_{i=1}^{\rr}h_{i}\left[\sum_{j=1}^{\rr}\lambda_{1ij}\frac{D^{\mu}n_{j}}{n}+\lambda_{2i}\frac{D^{\mu}T}{T}+\lambda_{4i}\frac{q}{h}E^{\mu}+\ell_{3i}\left(a^{\mu}+\frac{1}{nh}D^{\mu}p-\frac{q}{h}E^{\mu}\right)\right],
\label{Eq:hatQ1-A}
\end{align}
\end{subequations}
where we have defined 
\begin{align}
\hat{\lambda}_{1ij}:=\lambda_{1ij}-\frac{n_{i}}{n}\sum\limits_{k=1}^{\ell}\lambda_{1kj},\qquad\hat{\lambda}_{2i}:=\lambda_{2i}-\frac{n_{i}}{n}\sum\limits_{k=1}^{\ell}\lambda_{2k},\qquad\hat{\lambda}_{4i}:=\lambda_{4i}-\frac{n_{i}}{n}\sum\limits_{k=1}^{\ell}\lambda_{4k},
\end{align}
in order to simplify the expression for $\hat{\mathcal{J}}_{i}^{\mu}$. Notice additionally that the quantities $f$ and $\ell$ defined in Eqs.~(\ref{Eq:flDef}) 
are not invariant under a change of representation. Consequently, the frame-invariant fluxes transform according to
\begin{subequations}
\label{Eq:hatTransf}
\begin{align}
\hat{\mathrm{P}}\mapsto\hat{\mathrm{P}}+\sum_{j=1}^{\rr}\Xi_{0j}\left(\frac{\dot{n}_{j}}{n_{j}}+\theta\right)+\Xi_{1}\left(\frac{\dot{T}}{T}+\frac{k_{B}}{c_{v}}\theta\right),
\label{Eq:hatP1-1}\\
\hat{\mathcal{J}}_{i}^{\mu}\mapsto\hat{\mathcal{J}}_{i}^{\mu}+\Xi_{2i}\left(a^{\mu}+\frac{1}{nh}D^{\mu}p-\frac{q}{h}E^{\mu}\right),
\label{Eq:hatJ1-1}\\
\hat{\mathcal{Q}}^{\mu}\mapsto\hat{\mathcal{Q}}^{\mu}+\Xi_{3}\left(a^{\mu}+\frac{1}{nh}D^{\mu}p-\frac{q}{h}E^{\mu}\right),
\label{Eq:hatQ1-1}
\end{align}
\end{subequations}
where
\begin{subequations}
\begin{align}
\Xi_{0j} & :=\Omega_{\pi j}-\frac{k_{B}}{c_{v}}\Omega_{\epsilon j}+\sum_{i=1}^{\rr}\left(\frac{k_{B}}{c_{v}}e_{i}-k_{B}T\right)\Omega_{\nu ij}, & \Xi_{1} & :=\chi_{\pi}-\frac{k_{B}}{c_{v}}\chi_{\epsilon}+\sum_{i=1}^{\rr}\left(\frac{k_{B}}{c_{v}}e_{i}-k_{B}T\right)\chi_{\nu i},
\\[1ex]
\Xi_{2i} & :=\frac{k_{B}T}{h}\left(\xi_{\mathcal{J}i}-\frac{n_{i}}{n}\sum_{k=1}^{\rr}\xi_{\mathcal{J}k}\right), & \Xi_{3} & :=\frac{k_{B}T}{h}\left(\sum_{i=1}^{\rr}\xi_{\mathcal{J}i}-\frac{1}{h}\xi_{Q}\right).
\end{align}
\end{subequations}
It should be noted that the additional terms included in Eqs.~(\ref{Eq:hatTransf}) are contributions of order $\mathcal{O}(\partial^{2})$.

\subsection{Examples of frames}\label{SubSec:Examples of Frames}

As discussed in subsection~\ref{SubSec:First-order frame transformations} one has the freedom to choose any specific frame to describe the system. This amounts to assigning values to the coefficients $\alpha$'s, $\beta$'s and $\mu$'s in Eqs.~\eqref{Eq:deltatransf}. In practice, this is achieved by imposing matching conditions which match suitable components of $J^{\mu}_{i}$ and $T^{\mu\nu}$ to their equilibrium values. Within the whole set of possible frames, a particular family of them, which we refer to as particle frames, is defined through the condition
\begin{equation}
\sum_{i=1}^{\rr}J_{i}^{\mu}=nu^{\mu},
\end{equation}
which implies that the total particle current matches its zeroth-order contribution. This implies $\sum\limits_{i=1}^{\rr}\mathcal{J}_{i}^{\mu}=0$ and $\sum\limits_{i=1}^{\rr}\mathcal{N}_{i}=n$ which impose restrictions on the coefficients $\nu_{1ij}$, $\nu_{Ii}$, $\mu_{1j}$ and $\mu_{J}$, and fixes $d+1$ out of the $\rr + d + 1$ matching conditions for the system. Another relevant family corresponds to the energy frames, characterized by the condition
\begin{equation}
    u_{\mu}T^{\mu\nu}=-neu^{\nu},
\end{equation}
which also fixes $d+1$ matching conditions. More specifically, these frames set $\mathcal{E}=ne$ and $\mathcal{Q}^{\mu}=0$. Below we briefly describe three examples of frames within these families. 
\begin{enumerate}
\item Eckart frame for mixtures:\\
The natural generalization of the Eckart frame \cite{Eckart1940}
for mixtures is given by a particle frame in which the internal energy density and the individual
particle densities are fixed. Specifically
one imposes\footnote{Notice that these conditions, together with the general ones characterizing the particle frames defined above, add up to $\rr+d+1$ independent conditions and completely
determine the coefficients $\alpha$'s, $\beta$'s and $\mu$'s.}
\begin{equation}
\mathcal{E}=ne,\quad\mathcal{N}_{i}=n_{i}, \quad \text{for} \quad i = 1,\dots,\rr.
\end{equation}
\item Landau frame for mixtures:\\
From all the possible energy frames, the one that corresponds to the
generalization to mixtures of Landau's frame \cite{LandauLifshitz-Book6} is given by the additional conditions $\mathcal{N}_{i}=n_{i}$ for
$i=1,...,\rr$.
\item Family of trace-fixed particle frames for mixtures:\\
Within the particle frames, one can also introduce a matching condition on the trace of the stress-energy tensor by imposing
\begin{equation}
    \mathcal{E}-d\mathcal{P}=ne-dp \label{eq:matchTF}.
    \end{equation}
This condition, for the single species case, completely determines the frame and coincides with the frame considered in Refs.~\cite{SGS2025A,SGS2025B}. In the multiple species scenario here considered, condition~(\ref{eq:matchTF}) leads to a family of frames, leaving $\rr-1$ matching conditions still undetermined.
\end{enumerate}

By adopting the Eckart frame, we can provide the physical interpretation of the invariant coefficient $\zeta$ since, in this frame the quantity $\hat{\mathrm{P}}$ from Eq.~(\ref{Eq:Scalar Invariant-1}) reduces to $\hat{\mathrm{P}}=\mathcal{P}-p$, which according to Eq.~(\ref{Eq:Phat}) leads to $\mathcal{P}-p=-\zeta\theta$. It is clear from this that $\zeta$ corresponds to the bulk viscosity. Moreover, it is clear from Eq.~\eqref{Eq:ConstitutiveT} that the invariant transport coefficient $\eta$ is the shear viscosity. The physical interpretation of the remaining invariants, which are related to diffusive and thermal processes, is more involved. It will be clarified in section~\ref{Sec: Binary Mixture}, after the entropy production term has been analyzed.

\section{Entropy current and entropy production}\label{Section: Entropy Current}

In this section, we derive an expression for the entropy production in terms of the invariant fluxes obtained in the previous section, and we establish the conditions that must be imposed on the coefficients in order to ensure the fulfillment of the second law and Onsager's reciprocal relations. As a starting point, following~\cite{Israel1976,Israel1979,Israel1989}, we define the entropy current as: 
\begin{equation}
S^{\mu}:=\frac{1}{T}\left(pu^{\mu}-T^{\mu\nu}u_{\nu}-\sum_{i=1}^{\rr}\mu_{i}J_{i}^{\mu}\right),\label{Eq:EntropyFlux}
\end{equation}
where $\mu_{i}:=e_{i}-Ts_{i}+k_{B}T$ is the relativistic chemical potential associated with the $i$-th species, with $s_{i}$ denoting the corresponding entropy per particle. Note that the intensive quantities $\mu_i$ and $s_i$ are functions of the temperature parameter $T$ only; hence, according to the Gibbs relation, they satisfy (see appendix \ref{App:Thermo} for details) 
\begin{equation}
n_{i}d\mu_{i}=-n_{i}s_{i}dT+d\left(n_{i}k_{B}T\right).
\label{Ass_GD}
\end{equation}
The entropy production can be cast in the form: 
\begin{equation}
\begin{aligned}T\nabla_{\mu}S^{\mu} & =-\left(\mathcal{E}-ne\right)\frac{\dot{T}}{T}-(\mathcal{P}-p)\theta-\sum_{i=1}^{\rr}\left(\mathcal{N}_{i}-n_{i}\right)\left(\dot{\mu}_{i}-\mu_{i}\frac{\dot{T}}{T}\right)\\
 & \quad-\sum_{i=1}^{\rr}\mathcal{J}_{i}^{\mu}\left(D_{\mu}\mu_{i}-\mu_{i}\frac{D_{\mu}T}{T}-q_{i}E_{\mu}\right)-\mathcal{Q}^{\mu}\left(\frac{D_{\mu}T}{T}+a_{\mu}\right)-\mathcal{T}^{\mu\nu}\sigma_{\mu\nu},
\end{aligned}
\label{Eq:EPMixtures}
\end{equation}
where, to derive this result, we have used the decompositions given in Eqs.~\eqref{Eq:JTKovtun} together with Eqs.~\eqref{Eq:GibbsPotential}. In compact form, we write the entropy production as the following sum: 
\begin{equation}
T\nabla_{\mu}S^{\mu}=\sigma_{0}+\sigma_{1}+\sigma_{2} =: \sigma,\label{EntropyP:1}
\end{equation}
where $\sigma_{0}$, $\sigma_{1}$, and $\sigma_{2}$ represent the scalar, vector, and tensor contributions, respectively. We now express the scalar, vector, and tensor contributions for the entropy production in terms of the invariant fluxes, namely
$\hat{\mathrm{P}}$, $\hat{\mathcal{J}}_{i}^{\mu}$, $\hat{\mathcal{Q}}^{\mu}$, and $\mathcal{T}^{\mu \nu}$.

For the scalar part we notice that $\sigma_{0}$ can be expressed in terms of the invariant $\hat{\mathrm{P}}$,
defined in Eq.~(\ref{Eq:Scalar Invariant-1}), through the use of Eqs.~\eqref{Eq:EulerTrunc1-1},~\eqref{Eq:EulerTrunc2-1}, and \eqref{Eq:chemicap}, leading to: 
\begin{equation}
\sigma_{0}:=-\left(\mathcal{E}-ne\right)\frac{\dot{T}}{T}-(\mathcal{P}-p)\theta-\sum_{i=1}^{\rr}\left(\mathcal{N}_{i}-n_{i}\right)\left(\dot{\mu}_{i}-\mu_{i}\frac{\dot{T}}{T}\right)=-\hat{\mathrm{P}}\theta
 + \mathcal{O}\left(\partial^{3}\right),
\end{equation}
which, using Eq.~\eqref{Eq:Phat}, yields
\begin{equation}\label{Eq:EntropyInvariant1-1}
    \sigma_{0} = \zeta\theta^{2}+\mathcal{O}\left(\partial^{3}\right). 
\end{equation}

Next, we express the vector contribution in terms of the invariant fluxes defined in Eqs.~\eqref{Eq:VectorInvariants}. This yields the relation:
\begin{equation}
\sigma_{1} :=-\sum_{i=1}^{\rr}\mathcal{J}_{i}^{\mu}\left(D_{\mu}\mu_{i}-\mu_{i}\frac{D_{\mu}T}{T}-q_{i}E_{\mu}\right)-\mathcal{Q}^{\mu}\left(\frac{D_{\mu}T}{T}+a_{\mu}\right)=-\sum_{i=1}^{\rr}\hat{\mathcal{J}_{i}^{\mu}}B_{i\mu}-\hat{Q}^{\mu}A_{\mu} + \mathcal{O}\left(\partial^{3}\right),
\label{Eq:EntropyProductionVector}
\end{equation}
where we have introduced the invariant forces
\begin{align}\label{Eq:DiffusiveForce}
B_{i\mu} := k_B T D_\mu\left(\frac{\mu_i}{k_B T}\right) - q_i E_\mu + h_i A_{\mu},\qquad 
A_{\mu}:=\frac{D_{\mu}T}{T}+a_{\mu},
\end{align}
which correspond to the diffusive forces $B_{i\mu}$ and the generalized
thermal force $A_{\mu}$. Using Eq.~(\ref{Eq:chemicap}) together with Eq.~(\ref{Eq:EulerTrunc3-1}) one can rewrite the above expression as:
\begin{equation}\label{Eq:DiffusiveForceB}
B_{i\mu}=k_{B}T\left(\frac{D_{\mu}n_{i}}{n_{i}}-\frac{D_{\mu}n}{n}\right)+\left(h_{i}-h\right)a_{\mu}-\left(q_{i}-q\right)E_{\mu} + \mathcal{O}(\partial^{2}).
\end{equation}
It is interesting to notice at this point that the structure of the entropy production, when written in terms of the invariant fluxes, is such that the force conjugate to the diffusion flux does not contain the temperature gradient and the one conjugate to the heat flux does not depend on the gradient of the concentration. Achieving such a structure is what motivates the redefinition of the fluxes in Refs.~\cite{GrootMazur1983,DeGroot1980}, while in the present case it is a consequence of the frame-invariant definitions. 

Finally, the entropy production in the tensor block is given by:
\begin{equation}\label{Eq:EntropyProductionTensor}
    \sigma_{2} := -\mathcal{T}^{\mu \nu}\sigma_{\mu \nu} = 2\eta \sigma^{\mu \nu}\sigma_{\mu \nu} + \mathcal{O}(\partial^{3}).
\end{equation}

As outlined in Ref.~\cite{GrootMazur1983}, in accordance with the established tenets of linear irreversible thermodynamics, the
entropy production should adhere to the following structure: 
\begin{equation}\label{OnsagerRR-1}
        \sigma = -\mathbf{J} \cdot \mathbf{X} \quad \text{with} \quad\mathbf{J} = \mathbf{L} \mathbf{X},
\end{equation} 
where $\mathbf{J}$ and $\mathbf{X}$ represent the independent dissipative fluxes and the independent thermodynamic forces respectively, and $\mathbf{L}$ refers to the matrix formed by the phenomenological coefficients. Onsager's reciprocal relations are equivalent to the statement that the matrix $\mathbf{L}$ is symmetric and the second law of thermodynamics requires that its symmetric part be negative definite.

Before we proceed, we make the following observations. First, notice that Eq.~(\ref{OnsagerRR-1}) can be decoupled into scalar,  vector, and tensor blocks, as is evident from Eqs.~(\ref{Eq:EntropyInvariant1-1}), (\ref{Eq:EntropyProductionVector}), and (\ref{Eq:EntropyProductionTensor}). Since the scalar and tensor blocks just involve one transport coefficient each, Onsager's relations are automatically satisfied and the second law requires that the bulk and shear viscosity coefficients are positive: $\zeta > 0$ and $\eta > 0$. Second, we point out that the fluxes and forces in the vector block are not independent from each other, since they are subject to the algebraic constraints\footnote{For a discussion of this point see Ref.~\cite{GrootMazur1983} chapter 6, section 3.}
\begin{equation}\label{Eq:DependentFluxesForces}
\sum_{i=1}^{\rr}\hat{\mathcal{J}}_{i}^{\mu} = 0,\qquad
\sum_{i=1}^{\rr} n_i B_{i\mu} = 0.
\end{equation}
For this reason, in what follows, we rewrite the vector block in terms of independent fluxes and forces, such that the components of $\mathbf{J}$ and $\mathbf{X}$ in Eq.~\eqref{OnsagerRR-1} are independent, as required. From now on, for simplicity, we omit writing explicitly the higher-order terms of the form $\mathcal{O}(\partial^{2})$ in the constitutive relations and $\mathcal{O}(\partial^{3})$ in the entropy production.

\subsection{Reduced flux-force relations for the vector block}

In order to express the vector sector in terms of independent fluxes and forces  we rewrite the vector contribution to the entropy production in Eq.~\eqref{Eq:EntropyProductionVector} as
\begin{equation}\label{Eq:EntropyPjhatQhat}
    \sigma_{1} = - \sum_{i=1}^{\rr-1}\hat{j}^{\mu}_{i}\hat{b}_{i\mu} - \hat{\mathcal{Q}}^{\mu}A_{\mu}, 
\end{equation}
where we have defined
\begin{equation}\label{Eq:jkmu}
    \hat{j}^{\mu}_{k} := \sum_{i=1}^{k}\hat{\mathcal{J}}^{\mu}_{i}, \qquad 
    \hat{b}_{k}^{\mu} := B_{k}^{\mu} - B_{k+1}^{\mu}, \qquad k = 1,2,\dots, \rr-1.
\end{equation}
Due to the modified fluxes and forces appearing in $\sigma_{1}$, the constitutive relations must be adjusted accordingly. To achieve this, we first re-express Eqs.~(\ref{Eq:hatJ1-A}) and (\ref{Eq:hatQ1-A}) in terms of the invariant forces $B_{i}^{\mu}$ and $A^{\mu}$ defined in Eqs.~\eqref{Eq:DiffusiveForce} and~\eqref{Eq:DiffusiveForceB}. This results in the following equations:
\begin{subequations}
\begin{align}
\hat{\mathcal{J}}_{i}^{\mu} & =\sum_{j=1}^{\rr}\bar{\gamma}_{1ij}B_{j}^{\mu}+\bar{\gamma}_{2i}A^{\mu}+\bar{\gamma}_{3i}a^{\mu}+\bar{\gamma}_{4i}E^{\mu},\\
\hat{\mathcal{Q}}^{\mu} & =\sum_{j=1}^{\rr}\bar{\kappa}_{1j}B_{j}^{\mu}+\bar{\kappa}_{2}A^{\mu}+\bar{\kappa}_{3}a^{\mu}+\bar{\kappa}_{4}E^{\mu},
\end{align}
\end{subequations}
with coefficients 
\begin{equation}
\begin{aligned}
\bar{\gamma}_{1ij} & :=\frac{n_{j}}{p}\hat{\lambda}_{1ij}, & \qquad
\bar{\kappa}_{1j} & :=-\frac{n_{j}}{p}\sum_{i=1}^{\rr}h_{i}\lambda_{1ij},\\
\bar{\gamma}_{2i} & :=\hat{\lambda}_{2i}-\sum_{j=1}^{\rr}\frac{n_{j}}{n}\hat{\lambda}_{1ij}, & \bar{\kappa}_{2} & :=\sum_{i=1}^{\rr}h_{i}\left(\sum_{j=1}^{\rr}\frac{n_{j}}{n}\lambda_{1ij}-\lambda_{2i}\right),\\
\bar{\gamma}_{3i} & :=-\left(\hat{\lambda}_{2i}+\sum_{j=1}^{\rr}\frac{n_{j}e_{j}}{p}\hat{\lambda}_{1ij}\right), & \bar{\kappa}_{3} & :=\sum_{i=1}^{\rr}h_{i}\left(\sum_{j=1}^{\rr}\frac{n_{j}e_{j}}{p}\lambda_{1ij}+\lambda_{2i}\right),\\
\bar{\gamma}_{4i} & :=\frac{q}{h}\hat{\lambda}_{4i}+\sum_{j=1}^{\rr}\frac{n_{j}q_{j}}{p}\hat{\lambda}_{1ij}, & \bar{\kappa}_{4} & :=-\sum_{i=1}^{\rr}h_{i}\left(\sum_{j=1}^{\rr}\frac{n_{j}q_{j}}{p}\lambda_{1ij}+\frac{q}{h}\lambda_{4i}\right).
\end{aligned}
\label{Eq:Transport-coefficientsbar}
\end{equation}
The subsequent step entails the replacement of $ \hat{\mathcal{J}}^{\mu}_{i} $ by $ \hat{j}^{\mu}_{k} $ and $ B^{\mu}_{j} $ by $ \hat{b}^{\mu}_{j} $. To accomplish this, we recall the definition in Eq.~(\ref{Eq:jkmu}) and write
\begin{equation}\label{Eq:BkRecursive}
\begin{aligned}
    B_{k}^{\mu} & = \hat{b}_{k}^{\mu} + B_{k+1}^{\mu} \\
             & = \hat{b}_{k}^{\mu} + \hat{b}_{k+1}^{\mu} + \cdots
             + \hat{b}_{\rr-1}^{\mu} + B_{\rr}^{\mu},
\end{aligned}
\end{equation}
which, taking into account the constraint  $ \sum\limits_{k=1}^{\rr} n_{k}B_{k}^{\mu} = 0 $, leads to
\begin{equation}
    B_{\rr}^{\mu} = -\frac{1}{n}\sum_{j=1}^{\rr-1}\left(\sum_{k=1}^{j}n_{k}\right)\hat{b}_{j}^{\mu}.
\end{equation}
For the case $k < \rr$, Eq.~(\ref{Eq:BkRecursive}) can now be rewritten as
\begin{equation}\label{Eq:Bkmu}
    B_{k}^{\mu} = \sum_{j=k}^{\rr-1}\hat{b}_{j}^{\mu} - \frac{1}{n}\sum_{j=1}^{\rr-1}N_j\hat{b}_{j}^{\mu} 
    =
    \sum_{j=k}^{\rr-1}\frac{\bar{N}_{j}}{n}\hat{b}_{j}^{\mu} - \sum_{j=1}^{k-1}\frac{N_{j}}{n}\hat{b}_{j}^{\mu},
\end{equation}
with $ N_{j}:= \sum\limits_{i=1}^{j}n_{i} $ and $\bar{N}_{j} := \sum\limits_{i=j+1}^{\rr}n_{i}$. With this at our disposal, we are able to establish the constitutive relations for $(\hat{j}^{\mu}_{k}, \hat{\mathcal{Q}}^{\mu})$ in terms of $(\hat{b}^{\mu}_{j},A^\mu,a^\mu,E^\mu)$:
\begin{subequations}
\label{Eq:ConstitutivejQhat}
\begin{align}
\hat{j}^{\mu}_{k} & = \sum^{\rr-1}_{j=1}\gamma^{*}_{1kj}\hat{b}^{\mu}_{j} + \gamma^{*}_{2k}A^{\mu} + \gamma^{*}_{3k}a^{\mu} + \gamma^{*}_{4k}E^{\mu}, \qquad k = 1,2,\dots, \rr-1, \label{Eq:Constitutivejmuk}\\
\hat{\mathcal{Q}}^{\mu} & = \sum^{\rr-1}_{j=1}\kappa^{*}_{1j}\hat{b}^{\mu}_{j}+\bar{\kappa}_{2}A^{\mu}+\bar{\kappa}_{3}a^{\mu}+\bar{\kappa}_{4}E^{\mu}, \label{Eq:ConstitutivehatQmu_bmu}
\end{align}
\end{subequations}
where the coefficients are given by
\begin{subequations}
\begin{align}
   \gamma^{*}_{1kj} &:= \frac{\bar{N}_{j}}{n}\sum_{i=1}^{k}\sum_{p=1}^{j}\bar{\gamma}_{1ip} - \frac{N_{j}}{n}\sum_{i=1}^{k}\sum_{p=j+1}^{\rr}\bar{\gamma}_{1ip}, 
\quad 
\gamma^{*}_{2k} :=
\sum^{k}_{i=1}\bar{\gamma}_{2i},
\quad
\gamma^{*}_{3k}
:=
\sum^{k}_{i=1}\bar{\gamma}_{3i},
\quad
\gamma^{*}_{4k}
:=
\sum^{k}_{i=1}\bar{\gamma}_{4i}, \\
\kappa^{*}_{1j} & := 
\frac{\bar{N}_{j}}{n}
\sum_{p=1}^{j}\bar{\kappa}_{1p}
-
\frac{N_j}{n}
\sum_{p=j+1}^{\rr}\bar{\kappa}_{1p}.
\end{align}
\end{subequations}
In scenarios where thermoelectric effects are of interest, we introduce the invariant electric current $\hat{\mathcal{I}}^{\mu}$, defined as
\begin{equation}
\hat{\mathcal{I}}^{\mu} := \sum_{k=1}^{\rr}q_{k}\hat{\mathcal{J}}^{\mu}_{k}.
\label{eq:electricalcurrent}
\end{equation}
This can be reexpressed in terms of the independent fluxes in the following way:
\begin{equation}
\hat{\mathcal{I}}^{\mu} =  \sum_{k=1}^{\rr-1}\left(q_{k} - q_{k+1}\right)\hat{j}^{\mu}_{k}.
\label{eq:electricalcurrent}
\end{equation}

\subsection{Second law and Onsager's relations for the vector block}\label{subsec:vectorblock}

We now turn to the conditions that must be imposed on the vector sector
in order to express $\sigma_{1}$ as a positive-definite quadratic form and thereby
ensure consistency with the second law.
To this end, we first write the constitutive equations given in Eqs.~(\ref{Eq:ConstitutivejQhat}) in matrix form as:
\begin{equation}
\mathbf{J}_\mathrm{v} = \mathbf{L}_\mathrm{v}\mathbf{X}_\mathrm{v},
\label{Eq:ConstititiveEqsMatrix}
\end{equation}
with the vectors
\begin{equation}
\mathbf{J}_\mathrm{v}^{\mathrm{T}} = \begin{pmatrix}
    \hat{j}_{1}^{\mu}, & \dots, & \hat{j}_{\rr-1}^{\mu}, & \hat{\mathcal{Q}}^{\mu}
\end{pmatrix}, 
\qquad
\boldsymbol{\mathrm{X}}^{\mathrm{T}}_\mathrm{v} =
\begin{pmatrix}
    \hat{b}_{1}^{\mu}, & \dots, & \hat{b}_{\rr-1}^{\mu}, & A^{\mu}, & a^{\mu}, & E^{\mu} 
\end{pmatrix},
\end{equation}
where $\mathrm{T}$ denotes transposition and  $\mathbf{L}_\mathrm{v}$ is an $\rr \times \left(\rr+2\right)$ matrix. Next, we split the contributions to the constitutive equations in order to isolate the terms that involve $\hat{b}_{i}^{\mu}$ and $A^\mu$ as follows:
\begin{equation}
\mathbf{J}_\mathrm{v} = \mathbf{L}_{1}\mathbf{X}_{1}+\mathbf{L}_{2}\mathbf{X}_{2},
\end{equation}
where $\mathbf{L}_{1}$ and $\mathbf{L}_{2}$ are given by
\begin{equation}
\mathbf{L}_{1} := 
\begin{pmatrix}\gamma_{1,11}^{*} & \gamma_{1,12}^{*} & \cdots & \gamma_{1,1(\rr-1)}^{*} & 
\gamma_{2,1}\\
\gamma_{1,21}^{*} & 
\gamma_{1,22}^{*} & \cdots & \gamma_{1,2(\rr-1)}^{*} & 
\gamma_{2,2}\\
\vdots & \vdots & \ddots & \vdots & \vdots\\
\gamma_{1,(\rr-1)1}^{*} & \gamma_{1,(\rr-1)2}^{*} & \cdots & \gamma_{1,(\rr-1)(\rr-1)}^{*} & \gamma_{2,(\rr-1)}\\
\kappa_{1,1}^{*} & 
\kappa_{1,2}^{*} & \cdots & \kappa_{1,\rr-1}^{*} & 
\bar{\kappa}_{2}
\end{pmatrix},\qquad
\mathbf{L}_{2} := \begin{pmatrix}
\gamma^{*}_{3,1} & \gamma^{*}_{4,1}\\
\gamma^{*}_{3,2} & \gamma^{*}_{4,2}\\
\vdots & \vdots\\
\gamma^{*}_{3,\rr-1} & \gamma^{*}_{4,\rr-1}\\
\bar{\kappa}_{3} & \bar{\kappa}_{4}
\end{pmatrix},
\end{equation}
and 
\begin{equation}
\mathbf{X}_{1}^{\mathrm{T}} := 
\begin{pmatrix}
    \hat{b}_{1}^{\mu}, & \dots, & \hat{b}_{(\rr-1)}^{\mu}, & A^{\mu}
\end{pmatrix},
\qquad
\mathbf{X}_{2}^{\mathrm{T}} := 
\begin{pmatrix}
    a^{\mu}, & E^{\mu}
\end{pmatrix}.
\end{equation}
In terms of these quantities the vector contribution to the entropy production $\sigma_1$ can be written as
\begin{equation}
\sigma_{1}=-\mathbf{X}_{1}^{\mathrm{T}}\mathbf{J}_\mathrm{v} = -\mathbf{X}_{1}^{\mathrm{T}}\left(\mathbf{L}_{1}\mathbf{X}_{1}+\mathbf{L}_{2}\mathbf{X}_{2}\right)=-\mathbf{X}_{1}^{\mathrm{T}}\mathbf{L}_{1}\mathbf{X}_{1}-\mathbf{X}_{1}^{\mathrm{T}}\mathbf{L}_{2}\mathbf{X}_{2}.\label{Eq:EntropyMatrixVector1}
\end{equation}
Therefore, in order for the quadratic terms in $\sigma_1$ to be positive definite, one needs $\mathbf{L}_{2} = 0$ and the symmetric part of $\mathbf{L}_{1}$ to be negative definite, such that $\sigma_{1} = -\mathbf{X}_{1}^{\mathrm{T}}\mathbf{L}_{1}\mathbf{X}_{1} \geq 0$. The requirement $\boldsymbol{\mathrm{L}}_{2} = 0$ translates into the following set of constraints:
\begin{equation}
    \bar{\gamma}_{3i} = 0, \quad
    \bar{\gamma}_{4i} = 0,  \quad
    \bar{\kappa}_{3} = 0, \quad 
    \bar{\kappa}_{4} = 0,
\qquad i = 1,2\dots,\rr-1,
\end{equation}
which leads to the relations
\begin{subequations}
\label{Eq:rest_termo}
\begin{alignat}{2}
\hat{\lambda}_{2i}
&= -\sum_{j=1}^{\rr}\frac{n_{j} e_{j}}{p}\hat{\lambda}_{1ij},
\qquad&
\sum_{i=1}^{\rr}h_{i}\lambda_{2i}
&= -\sum_{i,j=1}^{\rr}h_{i}\frac{n_{j} e_{j}}{p}\lambda_{1ij},
\label{eq:rest_termo1}\\
\hat{\lambda}_{4i}
&= -\frac{h}{q}\sum_{j=1}^{\rr}\frac{n_{j}q_{j}}{p}\hat{\lambda}_{1ij},
\qquad&
\sum_{i=1}^{\rr}h_{i}\lambda_{4i}
&= -\frac{h}{q}\sum_{i,j=1}^{\rr}h_{i}\frac{n_{j}q_{j}}{p}\lambda_{1ij},
\label{eq:rest_termo2}
\end{alignat}
\end{subequations}
and further reduces the number of invariants coefficients in the vector sector from $\rr^{2} + 2\rr$ to $\rr^{2}$. Finally, the constitutive equations in the vector sector that are consistent with the second law of thermodynamics are given by
\begin{subequations}
\label{Eq:JQRes}
\begin{align}
\hat{j}^{\mu}_{i} &=\sum^{\rr-1}_{j=1}\gamma^{*}_{1ij}\hat{b}^{\mu}_{j}+\gamma^{*}_{2i}A^{\mu},\qquad i=1,2,\ldots,\rr-1,
\label{Eq:Jota1Res} \\
\hat{\mathcal{Q}}^{\mu} & = \sum^{\rr-1}_{j=1}\kappa^{*}_{1j}\hat{b}^{\mu}_{j}+\bar{\kappa}_{2}A^{\mu}, \label{Eq:Qu1Res}
\end{align}
\end{subequations}
with
\begin{subequations}
\begin{alignat}{2}
    \gamma^{*}_{1kj} 
    &=
    \frac{1}{np}\sum_{i=1}^{k}\sum_{p=1}^{j}\sum_{l=j+1}^{\rr}
    n_{p}n_{l}
    \left(\hat{\lambda}_{1ip} - \hat{\lambda}_{1il}\right),
    \qquad
    &
    \kappa^{*}_{1j} 
    &=
    -\frac{1}{np}
    \sum_{p=1}^{j} \sum_{l=j+1}^{\rr}
    n_{p}n_{l}
    \sum_{i=1}^{\rr}h_i
    \left(\lambda_{1ip} - \lambda_{1il} \right),
    \\[1ex]
    \gamma^{*}_{2k} 
    &=
    -\sum_{i=1}^{k}\sum_{j=1}^{\rr}
    h_{j}\frac{n_{j}}{p}\hat{\lambda}_{1ij},
    \qquad
    &
    \bar{\kappa}_{2}  
    &=
    \sum_{i,j=1}^{\rr}
    h_{i}\left(\frac{h_{j}n_{j}}{p}\lambda_{1ij}\right),
\end{alignat}
\end{subequations}
such that the symmetric part of $\mathbf{L}_{1}$ is negative-definite. Notice that if Onsager's relations $\mathbf{L}_{1}=\mathbf{L}_{1}^{\mathrm{T}}$ are imposed, the number of independent coefficients further reduces to $\rr(\rr+1)/2$.

In the next section we restrict the analysis to the binary mixture in order to associate the coefficients in the constitutive equations~\eqref{Eq:JQRes} with the transport coefficients describing direct and cross effects as defined in the literature.

\section{Binary mixture}\label{Sec: Binary Mixture}

From this point forward the analysis will be restricted to a two-component mixture, which corresponds to setting $\rr = 2$ in the formulae above. To simplify the notation, the two species are labeled by $a$ and $b$ instead of $1$ and $2$. In this case the particle currents $J_{i}^{\mu}$ and stress-energy tensor $T^{\mu\nu}$ are given as:
\begin{subequations}
\begin{eqnarray}
J_{a}^{\mu} & = & \mathcal{N}_{a}u^{\mu}+\mathcal{J}_{a}^{\mu},\\
J_{b}^{\mu} & = & \mathcal{N}_{b}u^{\mu}+\mathcal{J}_{b}^{\mu},\\
T^{\mu\nu} & = & \mathcal{E}u^{\mu}u^{\nu}+\mathcal{P}\Delta^{\mu\nu}+2u^{(\mu}\mathcal{Q}^{\nu)}+\mathcal{T}^{\mu\nu},
\end{eqnarray}
\end{subequations}
and they satisfy the balance equations 
\begin{align}
\nabla_{\mu}J_{a}^{\mu}=0,\qquad\nabla_{\mu}J_{b}^{\mu}=0,\qquad\nabla_{\mu}T^{\mu\nu}+I_{\mu}F^{\mu\nu}=0.
\end{align}
In this scenario, the constitutive equations~(\ref{Eq:Constitutive}) involve $32$ transport coefficients which reduce to the  $10$ frame- and representation-invariant quantities $\eta$, $\zeta$, $\lambda_{1aa}$, $\lambda_{1ab}$, $\lambda_{1ba},\lambda_{1bb},\lambda_{2a},\lambda_{2b},\lambda_{4a},\lambda_{4b}$.

Whereas the physical meaning of $\eta$ and $\zeta$ has been explained at the end of Section~\ref{SubSec:Examples of Frames}, it remains to elucidate the significance of the $8$ $\lambda$'s. It is expected that these coefficients are related to the thermal and electric conductivity, the diffusion coefficient, and the coefficients corresponding to the cross effects. The resulting flux-force relations allow us to establish the relativistic counterparts of the Fourier, Ohm, and Fick laws, as well as the Soret, Dufour, Benedicks, and Thompson effects. 

For the binary mixture, the constitutive equations reduce to
\begin{subequations}
\label{Eq:InvariantBinary}
\begin{align}
    \hat{j}^{\mu}_{a} & = L_{11} \hat{b}^{\mu}_{a} + L_{12}A^{\mu}, \label{Eq:Invariant Jmua Binary}\\
    \hat{\mathcal{Q}}^{\mu} & = L_{21}\hat{b}^{\mu}_{a} + L_{22}A^{\mu}, \label{Eq:Invariant Qmu Binary}
\end{align}
\end{subequations}
where
\begin{subequations}
\label{Eq:binarygammakappa}
\begin{align}
L_{11} &:=\gamma_{1aa}^{*}=  \frac{n_{a}n_{b}}{n^{2}p}\left[n_{b}\lambda_{1aa}-n_{a}\lambda_{1ba}-n_{b}\lambda_{1ab}+n_{a}\lambda_{1bb}\right],\\
L_{12} &:=\gamma_{2a}^{*}=  -\frac{1}{np}\left[h_{a}n_{a}\left(n_{b}\lambda_{1aa}-n_{a}\lambda_{1ba}\right)+h_{b}n_{b}\left(n_{b}\lambda_{1ab}-n_{a}\lambda_{1bb}\right)\right],\\
L_{21}&:=\kappa_{1a}^{*}=  -\frac{n_{a}n_{b}}{np}\left[h_{a}\left(\lambda_{1aa}-\lambda_{1ab}\right)+h_{b}\left(\lambda_{1ba}-\lambda_{1bb}\right)\right],\\
L_{22}&:=\bar{\kappa}_{2}=  \frac{1}{p}\left[h_{a}\left(h_{a}n_{a}\lambda_{1aa}+h_{b}n_{b}\lambda_{1ab}\right)+h_{b}\left(h_{a}n_{a}\lambda_{1ba}+h_{b}n_{b}\lambda_{1bb}\right)\right].
\end{align}
\end{subequations} 
The enforcement of the second law of thermodynamics leads to the following conditions:
\begin{equation}
L_{11}\leq0,\quad L_{22}\leq0,\quad L_{11}L_{22}- \frac{1}{4}(L_{12}+L_{21})^2\geq 0.
\end{equation}
In turn, the imposition of Onsager's reciprocal relations, i.e., $L_{12} = L_{21}$, gives rise to the additional constraint
\begin{eqnarray}
n_{a}\lambda_{1ba}-n_{b}\lambda_{1ab}=0,
\label{Eq:ConstraiLambdas}
\end{eqnarray}
which, when introduced in Eqs.~\eqref{Eq:binarygammakappa}, yields the following expressions for the transport coefficients:
\begin{subequations}
\begin{align}
L_{11} & = \frac{n_{a}n_{b}}{n^{2}p}\left[n_{b}\lambda_{1aa}-2n_{b}\lambda_{1ab}+n_{a}\lambda_{1bb}\right],\label{Eq:M11-1}\\
L_{12} = L_{21} & = -\frac{n_{b}}{np}\left[h_{a}n_{a}\lambda_{1aa}-\left(h_{a}n_{a}-h_{b}n_{b}\right)\lambda_{1ab}-h_{b}n_{a}\lambda_{1bb}\right],\label{Eq:M12-1}\\
L_{22} & =\frac{1}{p}\left[n_{a}h_{a}^{2}\lambda_{1aa}+2h_{a}h_{b}n_{b}\lambda_{1ab}+n_{b}h_{b}^{2}\lambda_{1bb}\right]\label{Eq:M22-1}.
\end{align}
\end{subequations}
In the following subsections, the transport coefficients $L_{ij}$ are identified with their non-relativistic counterparts for thermodiffusion and thermoelectric processes.

\subsection{Physical meaning of the coefficients $L_{11}$, $L_{12}$, $L_{21}$, and $L_{22}$}
\label{Appendix:Physical Coefficients}

In order to ascribe physical meaning to the coefficients $L_{ij}$ we rely on the non-relativistic constitutive relations formulated in Ref.~\cite{GrootMazur1983}. Accordingly, we rewrite Eqs.~(\ref{Eq:InvariantBinary}) in terms of the gradient of the temperature and the concentration, defined as $x_{i} := n_{i}/n$. Using
\begin{equation}
    \hat{b}^{\mu}_{a} = \frac{k_{B}T}{x_{b}}\frac{D^{\mu}x_{a}}{x_{a}} - \left(h_{a} - h_{b}\right)a^{\mu} - \left(q_{b} - q_{a}\right)E^{\mu},
\end{equation}
and assuming $E^{\mu} = 0 $ and $ p = const.$, it follows from Euler's equation that $a^{\mu} = 0$, which leads to the following constitutive equations
\begin{subequations}
\begin{align}
    \hat{j}^{\mu}_{a} & = L_{11} \frac{k_{B}T}{x_{b}} \frac{D^{\mu}x_{a}}{x_{a}} + L_{12} \frac{D^{\mu}T}{T}, \\
    \hat{\mathcal{Q}}^{\mu} & = L_{21} \frac{k_{B}T}{x_{b}} \frac{D^{\mu}x_{a}}{x_{a}}  + L_{22}  \frac{D^{\mu}T}{T}.
\end{align}
\end{subequations}
Comparison with Eqs.~(218) and (219) on page~275 in Ref.~\cite{DeGroot1980}, yields the following identifications:
\begin{equation}
    L_{22} := -\kappa T, \qquad
    L_{12} := -nx_{a}x_{b} T \mathcal{D}^{'}_{T}, \qquad
    L_{21} := -nx_{a}x_{b} T \mathcal{D}_{T}, \qquad
    L_{11} := -\frac{n x_{a}x_{b}}{k_{B}T}\mathcal{D}_{F},
\end{equation}
with $\kappa$, $ \mathcal{D}^{'}_{T} $, $\mathcal{D}_{T}$, and $ \mathcal{D}_{F} $ the Fourier, Dufour, Soret, and Fick coefficients, respectively, where Onsager's relations imply that $\mathcal{D}^{'}_{T} = \mathcal{D}_{T}$. 

\subsection{Ohm's law and electrothermal-thermoelectric effects}

An alternative interpretation of the transport coefficients is provided by restricting ourselves to electrical and chemical effects. In order to assess the corresponding constitutive equations, we again rely on Ref.~\cite{GrootMazur1983} and consider as dissipative fluxes the electrical current $\hat{\mathcal{I}}^{\mu}$ as given in Eq.~(\ref{eq:electricalcurrent}) and the spatial entropy flux, which can be written as
\begin{equation}
\hat{\mathcal{S}}^{\mu}=\frac{1}{T}\hat{\mathcal{Q}}^{\mu}+\sum_{k=1}^{\rr-1}s_{k}\hat{j}_{k}^{\mu},
\end{equation}
where $s_{k}=\frac{1}{T}\left(h_{k}-\mu_{k}\right)$ is the entropy per particle for the $k$'th species. For a binary mixture these fluxes are given by
\begin{equation}
\mathbf{J}'_\mathrm{v}:=\left(\begin{array}{c}
\hat{\mathcal{I}}^{\mu}\\
\hat{\mathcal{S}}^{\mu}
\end{array}\right)=\left(\begin{array}{c}
\left(q_{a}-q_{b}\right)\hat{j}_{a}^{\mu}\\
\frac{\hat{Q}^{\mu}}{T}+\left(s_{a}-s_{b}\right)\hat{j}_{a}^{\mu}
\end{array}\right),
\label{Eq:Jprime}
\end{equation}
and one seeks a flux-force relation of the form $\mathbf{J}'_\mathrm{v}=\mathbf{L}'_\mathrm{v}\mathbf{X}'_\mathrm{v}$ in which $\mathbf{L}'_\mathrm{v}$ retains the symmetry properties of the matrix $\mathbf{L}_\mathrm{v}$ involved in the constitutive equations obtained in Section~\ref{subsec:vectorblock}, which for a binary mixture reduce to $\mathbf{J}_\mathrm{v}=\mathbf{L}_\mathrm{v}\mathbf{X}_\mathrm{v}$ with
\begin{equation}
\mathbf{J}_\mathrm{v} :=\left(\begin{array}{c}
\hat{j}_{a}^{\mu}\\
\hat{Q}^{\mu}
\end{array}\right),\quad
\mathbf{X}_\mathrm{v} :=\left(\begin{array}{c}
\hat{b}_{a}^{\mu}\\
A^{\mu}
\end{array}\right),\quad
\mathbf{L}_\mathrm{v}:=\left(\begin{array}{cc}
L_{11} & L_{12}\\
L_{21} & L_{22}
\end{array}\right).
\label{Eq:JXL}
\end{equation}
It is straightforward to verify that in order to achieve the sought properties of $\mathbf{L}'_\mathrm{v}$, the corresponding transformation should have the form $\mathbf{J}'_\mathrm{v} = \mathbf{T}^\mathrm{T}\mathbf{J}_\mathrm{v}$, $\mathbf{X}'_\mathrm{v} = \mathbf{T}^{-1}\mathbf{X}_\mathrm{v}$, such that $\mathbf{L}'_\mathrm{v} = \mathbf{T}^\mathrm{T}\mathbf{L}_\mathrm{v}\mathbf{T}$ is symmetric whenever $\mathbf{L}_\mathrm{v}$ is. From Eqs.~\eqref{Eq:Jprime} and~\eqref{Eq:JXL} it follows that
\begin{equation}
\mathbf{T}=\left(\begin{array}{cc}
q_{a}-q_{b} & s_{a}-s_{b} \\
0 & \frac{1}{T}
\end{array}\right).
\end{equation}
such that the new forces $\mathbf{X}'_\mathrm{v} = \left( 
\bar{G}^{\mu}, \bar{A}^{\mu} \right)^{\mathrm{T}}$ are given by
\begin{subequations}
\begin{align}
\bar{G}^{\nu} & :=\frac{D^{\nu}(\mu_a-\mu_b)}{q_{a}-q_{b}}-E^{\nu}+\frac{\mu_a-\mu_b}{q_{a}-q_{b}}a^{\nu},\\
\bar{A}^{\nu} & :=D^{\nu}T+Ta^{\nu}.
\end{align}
\end{subequations}
This leads to the following constitutive equations
\begin{subequations}
\label{Eq:ElectroThermoConstRel}
\begin{align}
\hat{\mathcal{I}}^{\mu} & =L'_{11}\bar{G}^{\mu}+L'_{12}\bar{A}^{\mu},\\
\hat{\mathcal{S}}^{\mu} & =L'_{21}\bar{G}^{\mu}+L'_{22}\bar{A}^{\mu},
\end{align}
\end{subequations}
where
\begin{subequations}
\begin{align}
L'_{11} & =\left(q_{a}-q_{b}\right)^{2}L_{11},\\
L'_{12} & =\left(q_{a}-q_{b}\right)\left[L_{11}\left(s_{a}-s_{b}\right)+\frac{L_{12}}{T}\right],\\
L'_{21} & =\left(q_{a}-q_{b}\right)\left[L_{11}\left(s_{a}-s_{b}\right)+\frac{L_{21}}{T}\right],\\
L'_{22} & = \left(s_{a}-s_{b}\right)^2 L_{11} + \frac{s_a-s_b}{T}(L_{12} + L_{21}) + \frac{L_{22}}{T^2}.
\end{align}
\end{subequations}
Onsager's reciprocal relations clearly hold if the symmetry $L_{12}=L_{21}$ is imposed. From Eq.~\eqref{Eq:ElectroThermoConstRel} one can identify $L'_{11}$ with the electrical conductivity, while $L'_{12}$ and $L'_{21}$ are the coefficients corresponding to the electro-thermal and thermoelectric effects which are related to the Seebeck and Peltier coefficients in the non-relativistic
scenario \cite{GrootMazur1983}.

\section{Binary mixture in a family of trace-fixed particle frames}\label{Sec:Binary Mixture in the TFP-Frame}

Although the invariant expressions derived previously are general and important from a conceptual point of view, performing the time evolution for the relevant variables requires choosing a specific frame to obtain a closed system of PDE's. In this section we explicitly derive the set of hydrodynamic equations for a binary mixture within the family of TFP frames. We start by recalling the definition of the TFP frames given in Section~\ref{SubSec:Examples of Frames} in which the matching conditions
\begin{equation}
\mathcal{N}_{a} + \mathcal{N}_{b} = n_{a} + n_{b}, \qquad
\mathcal{E}-d\mathcal{P}=ne-dp, \qquad\mathcal{J}_{a}^{\mu}=-\mathcal{J}_{b}^{\mu},
\label{Eq:FTFP Conditions}
\end{equation}
are imposed. These conditions lead to the following form for the particle currents and stress-energy tensor
\begin{subequations}
\begin{eqnarray}
J_{a}^{\mu} & = & \mathcal{N}_{a}u^{\mu} + \mathcal{J}^{\mu}_{a},\\
J_{b}^{\mu} & = & \mathcal{N}_{b}u^{\mu} - \mathcal{J}^{\mu}_{a},\\
T^{\mu\nu} & = & \left(ne + \epsilon\right)u^{\mu}u^{\nu}+\left(p + \frac{\epsilon}{d}\right)\Delta^{\mu\nu}+2u^{(\mu}\mathcal{Q}^{\nu)} - 2\eta\sigma^{\mu \nu},
\end{eqnarray}
\end{subequations}
with constitutive equations given as:
\begin{subequations}
\label{eq:constTFP}
\begin{align}
\mathcal{N}_{a} = & \hspace{0.1cm} n_{a} + \tau\left(\dot{n}_{a} + n_{a}\theta\right),
\label{eq:Ncala}\\
\mathcal{N}_{b} = & \hspace{0.1cm} n_{b} - \tau\left(\dot{n}_{a} + n_{a}\theta\right),
\label{eq:Ncalb}\\
\epsilon = & -\frac{\zeta}{\left(\frac{k_{B}}{c_{v}}-\frac{1}{d}\right)^{2}}\left[\frac{\dot{T}}{T}+\frac{1}{d}\theta-\Gamma_{0}\left(\frac{\dot{n}_{a}}{n_{a}}+\theta\right)-\Gamma_{1}\left(\frac{\dot{T}}{T}+\frac{k_{B}}{c_{v}}\theta\right)\right],
\label{eq:epsilon}\\
\mathcal{J}_{a}^{\mu} = &-\frac{n_{a}n_{b}}{n}\left\{ \mathcal{D}_{F}\left[\frac{n_b}{n_a}D^\mu\left( \frac{n_a}{n_b} \right) + \frac{h_{a}-h_{b}}{k_{B}T}a^{\mu}-\frac{q_{a}-q_{b}}{k_{B}T}E^{\mu}-\Gamma_{2a}\left(a^{\mu}+\frac{D^{\mu}p}{nh}-\frac{q}{h}E^{\mu}\right)\right]+T\mathcal{D}_{T}^{'}\left[\frac{D^{\mu}T}{T}+a^{\mu}\right]\right\},
\\
\mathcal{Q}^{\mu}= & -\Biggl\{ S_{1}\left[\frac{D^{\mu}T}{T}+a^{\mu}-\Gamma_{2}\left(a^{\mu}+\frac{D^{\mu}p}{nh}-\frac{q}{h}E^{\mu}\right)\right]+S_{2}\left[
\frac{n_b}{n_a}D^\mu\left( \frac{n_a}{n_b} \right) + \frac{h_{a} -h_{b}}{k_{B}T}a^{\mu} - \frac{q_{a}-q_{b}}{k_B T} E^{\mu} \right]\Biggr\},
\label{eq:Q}
\end{align}
\end{subequations}
where we have introduced the quantities
\begin{equation}
S_{1} := \kappa T + \frac{n_{a}n_{b}}{n}\left(h_{a} - h_{b}\right)  T\mathcal{D}_{T}^{'},\qquad S_{2} := \frac{n_{a}n_{b}}{n}k_{B}T\left(T\mathcal{D}_{T}+\frac{h_{a}-h_{b}}{k_{B}T}\mathcal{D}_{F}\right).
\end{equation}
Notice that we have here restored the freedom of representation discussed in Section~\ref{subsec:changeofrepresentation} within the constitutive equations. For the binary mixture in the TFP frame, one has two scalar and one vector combinations which are second order on shell. This freedom is encoded in the arbitrary coefficients $\tau$, $\Gamma_0$, $\Gamma_1$, $\Gamma_2$, and $\Gamma_{2a}$. It is important to point out that the structure in Eqs.~(\ref{eq:Ncala}) and (\ref{eq:Ncalb}) is compatible with the particle frame matching condition Eq.~(\ref{Eq:FTFP Conditions}). In particular, notice that the scalar combination $\dot{n}_a+n_a \theta$ is included in Eqs.~(\ref{eq:Ncala}) and (\ref{eq:Ncalb}). The reason for this will become clear once the evolution system is established below.

The balance laws together with the constitutive equations~\eqref{eq:constTFP} yield an evolution system for the quantities $(n_a,n_b,T,u^\mu,\epsilon,\mathcal{Q}^\mu)$. To write it down explicitly, it is convenient to replace $n_a$ and $n_b$ by the total particle number $n = n_a + n_b$ and the concentration $x_a := n_a/n$, which leads to
\begin{subequations}
\label{Eq:EvolSystem}
\begin{align}
\dot{n} & = -n\theta, 
\label{Eq:Evoln}\\
\tau \Ddot{x}_{a} & = \left(\mathcal{D}_{F} + M \beta_{0}\right)D^{\mu}D_{\mu}x_{a} - \left(L - M \frac{\beta_{2}}{T}\right)D^{\mu}D_{\mu}T + \frac{1}{n}\left[M\beta_{1} - x_{a}(1-x_{a})\Gamma_{2a}\frac{k_{B}T}{h}\mathcal{D}_{F}\right]D^{\mu}D_{\mu}n + f_{x_{a}},
\label{Eq:Evolxa}\\
\dot{T} & =\alpha_{4}\theta T+\alpha_{5}\epsilon T+f_{T},
\label{Eq:EvolT}\\
\dot{u}^{\mu} & = \beta_1\frac{D^{\mu}n}{n} + \beta_0 D^{\mu}x_{a} + \beta_{2}\frac{D^{\mu}T}{T}+\beta_{3}\mathcal{Q}^{\mu} + \beta_{4}E^{\mu}, 
\label{Eq:Evolu}\\
\dot{\epsilon} &  = -D_{\mu}Q^{\mu}+f_{\epsilon},
\label{Eq:Evolepsilon}\\
\dot{\mathcal{Q}}^{\mu} & = 2\eta D_{\nu}\sigma^{\mu\nu}-\frac{1}{d}D^{\mu}\epsilon + f_{\mathcal{Q}}^{\mu},
\label{Eq:EvolQ}
\end{align}
\end{subequations}
where the coefficients $L$, $M$, $\alpha_i$, $\beta_i$ and the lower-order terms $f_{x_a}$, $f_T$, $f_\epsilon$, $f^\mu_\mathcal{Q}$ are given by
\begin{subequations}
\begin{align}
L & := \frac{x_{a}\left(1-x_{a}\right)}{T}\left[\Gamma_{2a}\frac{k_{B}T}{h}\mathcal{D}_{F} - T\mathcal{D}^{'}_{T}\right], \\
M &:= x_{a}\left(1-x_{a}\right)\left[T \mathcal{D}^{'}_{T} + \mathcal{D}_{F}\left(\frac{h_{a} - h_{b}}{k_{B}T} - \Gamma_{2a}\right) \right],
\end{align}
\end{subequations}
\begin{subequations}
\begin{align}
\alpha_4 &:= -\frac{\Gamma_1\frac{k_B}{c_v}-\frac{1}{d}}{\Gamma_1-1},
&
\alpha_5 & := \frac{1}{\Gamma_1-1}
\left(\frac{k_B}{c_v}-\frac{1}{d}
\right)^2 \frac{1}{\zeta},
\label{eq:alpha_coefficients}
\\[1.5ex]
\beta_1 & := -
\frac{\Gamma_2\frac{k_BT}{h}S_1}{
S_1(\Gamma_2-1) -\frac{S_2}{k_BT}(h_a-h_b)
}, 
& 
\beta_0 &:= \frac{S_2}{x_a(1-x_a)
\left[S_1(\Gamma_2-1)
-\frac{S_2}{k_BT}(h_a-h_b)\right]},
\label{eq:beta_10}
\\[1.5ex]
\beta_2 & := \frac{S_1\left(
1-\Gamma_2\frac{k_BT}{h}
\right)}{
S_1(\Gamma_2-1)
-\frac{S_2}{k_BT}(h_a-h_b)
},
&
\beta_3
&:= \frac{1}{
S_1(\Gamma_2-1)
-\frac{S_2}{k_BT}(h_a-h_b)
},
\label{eq:beta_23}
\\[1.5ex]
\beta_4 & := \frac{S_1\Gamma_2\frac{q}{h}
-\frac{S_2}{k_BT}(q_a-q_b)}{S_1(\Gamma_2-1)
-\frac{S_2}{k_BT}(h_a-h_b)}.
\label{eq:beta_4}
\end{align}
\end{subequations}
and

\begin{subequations}
\begin{align}
f_{x_a} := {} &
-\left(1+\dot{\tau}\right)\dot{x}_a
-\frac{1}{n}a_\mu\mathcal{J}_a^\mu
+\frac{\mathcal{D}_F}{n}
\left(D^\mu x_a\right)\left(D_\mu n\right)
\nonumber\\
&\quad
+\frac{1}{n}a^\mu D_\mu\left[
nx_a(1-x_a)
\left(
T\mathcal{D}_T'
+\mathcal{D}_F
\left(\frac{h_a-h_b}{k_BT}-\Gamma_{2a}
\right)\right)\right]
\nonumber\\
&\quad
-\frac{1}{n}D_\mu \left[nx_a(1-x_a)
\left(\Gamma_{2a}\frac{k_B}{h}\mathcal{D}_F
-\mathcal{D}_T'\right)
\right]D^\mu T - \frac{D^\mu n}{n} D_\mu
\left[x_a(1-x_a)\Gamma_{2a}\frac{k_BT}{h}\mathcal{D}_F\right]
\nonumber\\
&\quad
-\frac{1}{n}D_\mu
\left[nx_a(1-x_a)\mathcal{D}_F\left(\frac{q_a-q_b}{k_BT} -\Gamma_{2a}\frac{q}{h}\right)E^\mu \right]
\nonumber\\
&\quad
-M\left[
D^\mu nD_\mu\left(\frac{\beta_1}{n}\right)
+D^\mu TD_\mu\left(\frac{\beta_2}{T}\right)
+D^\mu x_aD_\mu\beta_0
+\mathcal{Q}^\mu D_\mu\beta_3
+E^\mu D_\mu\beta_4
-\beta_3D_\mu\mathcal{Q}^\mu
-\beta_4D_\mu E^\mu
\right],
\\[1ex]
f_{\epsilon} := {}&
-\frac{d+1}{d}\theta\epsilon
+2\eta\,\sigma^{\mu\nu}\sigma_{\mu\nu}
-2a_\mu Q^\mu
+\frac{nTc_v}{\Gamma_1-1}\left(
\frac{k_B}{c_v}-\frac{1}{d}\right)
\left[\theta - \left(\frac{k_B}{c_v}-\frac{1}{d}\right)\frac{\epsilon}{\zeta}\right]
\nonumber\\
&\quad
+\left(q_a-q_b\right)\mathcal{J}_a^\mu E_\mu
-n\dot{x}_a(e_a-e_b)
-nc_vf_T,
\\[1ex]
f_T := {}&
-\frac{\Gamma_0}{\Gamma_1-1}
T\frac{\dot{x}_a}{x_a},
\\[1ex]
f_Q^\mu := {}&
2\sigma^{\mu\nu}
\left(T\frac{\partial\eta}{\partial T}
\frac{D_\nu T}{T}
+a_\nu\eta\right)
-\frac{d+1}{d}\left(
a^\mu\epsilon+\theta\mathcal{Q}^\mu
\right) - \left(\sigma^{\nu\mu} -\omega^{\mu\nu} \right)Q_\nu
\nonumber\\
&\quad
-nh\left[\left(\beta_1+\frac{k_BT}{h}
\right)\frac{D^\mu n}{n}
+\beta_0D^\mu x_a
+\left(\beta_2+\frac{k_BT}{h}
\right)\frac{D^\mu T}{T} + \beta_3\mathcal{Q}^\mu +\beta_4E^\mu
\right] - \Delta^\mu_{\ \beta}I_\alpha F^{\alpha\beta}.
\end{align}
\end{subequations}

Once a suitable equation for $e(T,x_a) = x_a e_a(T) + (1-x_a) e_b(T)$ and the diffusion coefficients $\eta$, $\zeta$, $\kappa$, $\mathcal{D}_F$, $\mathcal{D}_T$, and $\mathcal{D}'_F$ have been specified, equations~(\ref{Eq:EvolSystem}) provide a closed evolution system for the variables $(n,x_a,T,u^\mu,\epsilon,\mathcal{Q}^\mu)$. The hyperbolic structure, causality properties, and stability of this system will be analyzed in follow-up work. For now, we just emphasize two important points. First, note that the inclusion of the terms which are multiplied by $\tau$ in the right-hand sides of Eqs.~(\ref{eq:Ncala}) and (\ref{eq:Ncalb}) are necessary to achieve a hyperbolic evolution. Indeed, the operator acting on $x_a$ in Eq.~(\ref{Eq:Evolxa}) is parabolic if $\tau=0$, whereas it is a wave operator when $\tau > 0$ and $\mathcal{D}_F + M\beta_0 > 0$. The second point that we want to stress is that the parameters $\tau$, $\Gamma_1$, $\Gamma_2$, and $\Gamma_{2a}$ can still be chosen arbitrarily, which gives us hope that one can achieve a hyperbolic and causal evolution with stable global equilibria.

\section{Conclusion and final remarks}
\label{Section:Conclusions}

In this work, we developed a first-order formalism for dissipative relativistic gases composed of $\rr$ chemically non-reacting, charged species in the presence of an external electromagnetic field. We established general constitutive equations in the spirit of the BDNK-type formalisms including cross effects. The resulting relations are in alignment with the tenets of non-equilibrium thermodynamics while being invariant under changes of frame. Moreover, the transport coefficients appearing in these relations can be defined in such a way that they are both frame- and representation-invariant, and thus they posses a physical unambiguous interpretation.

The imposition of frame invariance naturally led to a redefinition of dissipative fluxes and their driving forces. While the tensor constitutive equation is automatically invariant and involves one transport coefficient, identified as the shear viscosity, the scalar sector reduces from $\rr+2$ non-equilibrium contributions to only one invariant dissipative flux. The corresponding transport coefficient is identified as the bulk viscosity. In the Eckart frame for mixtures, the definitions of the shear and bulk viscosities coincide with the ones in the standard literature \cite{DeGroot1980,CercignaniKremer-Book}. 

On the other hand, the vector sector is reduced to $\rr$ independent constitutive equations with $\rr^{2}$ invariant transport coefficients, which further reduce to $\rr(\rr+1)/2$ if Onsager's reciprocity relations are imposed. Remarkably, the invariant fluxes and forces in the vector sector coincide precisely with the ones adopted in standard non-equilibrium thermodynamics and kinetic theory \cite{GrootMazur1983,DeGroot1980,CercignaniKremer-Book} based on different arguments.

For the physical interpretation of the coefficients involved in the vector sector, the conditions for the second law of thermodynamics to hold were explicitly analyzed. For the particular case of a two-species fluid the transport coefficients associated with direct and cross-effects were identified. The corresponding constitutive equations were written both for the case of thermodiffusive effects, involving heat and diffusive fluxes, as well as for thermoelectric phenomena, involving the electrical current and the total entropy flux. In particular, we derived the general relativistic counterparts of the Fourier, Ohm, and Fick laws, as well as the Soret, Dufour, Benedicks, and Thompson effects. 

Finally, the explicit form of the system of transport equations for the binary mixture was explicitly written down in a frame belonging to the family of trace-fixed particle frames. The form in which the evolution equations are presented permits a detailed analysis of causality, hyperbolicity, and stability. Such study will be presented in a follow-up article.


\acknowledgments

We thank A. R. M\'endez for fruitful comments and discussions. This work was supported by SECIHTI under Grant No.~CBF-2025-G-1626. O.S. also acknowledges support from CIC grant No.~18315 to Universidad Michoacana de San Nicolás de Hidalgo. FS was supported by a SECIHTI postdoctoral fellowship.

\appendix

\section{Some relevant thermodynamic relations}
\label{App:Thermo}

In this appendix, we briefly recall the Euler and Gibbs-Duhem relations, which are useful for the entropy production calculation. To bridge the standard non-relativistic formulations with the relativistic theory, we consider a co-moving fluid element of volume $V$ evaluated in its local Lorentz frame, where spatial volumes and extensive quantities are locally well-defined. In this frame, for a fluid composed of several species, the Gibbs relation takes the standard form~\cite{Huang-Book}:
\begin{equation}
dU=TdS-pdV+\sum_{i=1}^{\rr}\mu_{i}dN_{i},
\label{Eq:GibbsR_U}
\end{equation}
with the extensive variables $(U,S,N_{i})$ representing the internal energy, entropy, and total number of particles for the $i$-th species, while the intensive variables $(T,p,\mu_{i})$ represents the temperature, pressure and chemical potential defined respectively as $T:=\left(\frac{\partial U}{\partial S}\right)_{V,N_{i}}$, $p:=\left(\frac{\partial U}{\partial V}\right)_{S,N_{i}}$, and $\mu_{i}:=\left(\frac{\partial U}{\partial N_{i}}\right)_{S,V,N_{j\neq i}}$. Due to the scaling property of the extensive variables, Eq.~(\ref{Eq:GibbsR_U}) implies the Euler relation
\begin{equation}
U=TS-pV+\sum_{i=1}^\rr \mu_{a}N_{i},
\end{equation}
and the relation for the intensive variables referred to as the Gibbs-Duhem relation 
\begin{equation}
SdT-Vdp+\sum_{i=1}^\rr N_{i}d\mu_{i}=0.
\end{equation}
The explicit dependence on $V$ disappears once we reformulate these expressions in terms of local densities. The density analog of the Gibbs, Euler, and Gibbs-Duhem relations are, respectively,
\begin{equation}
d\rho=Td\Bar{s}+\sum_{i=1}^{\rr}\mu_{i}dn_{i},
\qquad
ne=nTs-p+\sum_{i=1}^{\rr}\mu_{i}n_{i},
\qquad
\sum_{i=1}^{\rr} n_{i}d\mu_{i}=-nsdT+dp,
\label{Eq:GibbsPotential}
\end{equation}
where we have used the following notation $\rho=\frac{U}{V}=ne$, $\Bar{s}=\frac{S}{V}=ns$, and $n_{i}=\frac{N_{i}}{V}$, with $e$ and $s$ denoting the internal energy and entropy per total particle number, respectively. The chemical potential of the individual species is defined as
\begin{equation}
\mu_{i}=e_{i}-Ts_{i}+\frac{p_{i}}{n_{i}}, \qquad p_{i} := n_{i}k_{B}T,
\label{Eq:muiDef}
\end{equation}
which is compatible with the Euler relation, and satisfies
\begin{equation}
d\mu_{i}=-s_{i} dT + \frac{dp_{i}}{n_{i}}.
\end{equation}
From these equations we obtain the following useful expression:
\begin{equation}
T d\left(\frac{\mu_{i}}{T}\right)=-e_{i}\frac{dT}{T} + k_{B}T\frac{dn_{i}}{n_{i}}.
\label{Eq:chemicap}
\end{equation}



\bibliographystyle{ieeetr}
\bibliography{refsRelFluids}

\end{document}